\documentclass[12pt]{article}

\usepackage{amssymb}
\usepackage{amstext}
\usepackage{amsmath}

\def\w{\wedge}

\def\be{\begin{equation}}
\def\ee{\end{equation}}
\def\bea{\begin{eqnarray}}
\def\eea{\end{eqnarray}}

\begin{document}
\pagestyle{empty}
\begin{flushright}
\begin{tabular}{ll}
DAMTP-2007-54& \\
EMPG-07-10 & \\
ITFA-07-25 & \\
arXiv:0706.3119 [hep-th] & \\
05/07/07 & \\ [.3in]
\end{tabular}
\end{flushright}
\begin{center}
{\Large {\bf{$G_2$ Hitchin functionals at one loop}}} \\ [.35in]
{J. de Boer$\, {}^{1}$, {P. de Medeiros$\, {}^{2}$, S. El-Showk$\,
{}^{1}$ and A. Sinkovics$\,{}^{3}$}}
\\ [.2in]
$\,{}^{1}$\, {\emph{Institute for Theoretical Physics, University of Amsterdam, \\
Valckenierstraat 65, Amsterdam 1018 XE, The Netherlands.}} \\ [.2in]
$\,{}^{2}$\, {\emph{School of Mathematics and Maxwell Institute for Mathematical Sciences, \\
University of Edinburgh, King's Buildings, Mayfield Road, \\
Edinburgh EH9 3JZ, Scotland, U.K.}}  \\ [.2in]
$\,{}^{3}$\, {\emph{Department of Applied Mathematics and Theoretical Physics, \\
Centre for Mathematical Sciences, University of Cambridge, \\
Wilberforce Road, Cambridge CB3 0WA, U.K.}} \\ [.2in]
{\tt{jdeboer@science.uva.nl}}, {\tt{p.demedeiros@ed.ac.uk}}, {\tt{sheer@science.uva.nl}}, {\tt{a.sinkovics@damtp.cam.ac.uk}}
\\ [.45in]

{\large{\bf{Abstract}}} \\ [.15in]
\end{center}
We consider the quantization of the effective target space
description of topological M-theory in terms of the Hitchin
functional whose critical points describe seven-manifolds with
$G_2$ structure. The one-loop partition function for this theory
is calculated and an extended version of it, that is related to
generalized $G_2$ geometry, is compared with the topological $G_2$
string. We relate the reduction of the effective action for the
extended $G_2$ theory to the Hitchin functional description of the
topological string in six dimensions. The dependence of the
partition functions on the choice of background $G_2$ metric is
also determined.

\clearpage
\pagestyle{plain}
\pagenumbering{arabic}




\section{\large{Introduction}}

Topological string theory on Calabi-Yau manifolds has been the source of
many recent insights in the structure of gauge theories and black holes.
The traditional construction for topological strings is
in terms of topologically twisted worldsheet A- and B-models,
computing K\"{a}hler and complex structure deformations. The topological
information these theories compute is encoded in Gromov-Witten invariants.

More recently a target space quantum foam reformulation of the A-model in
terms of the K\"{a}hler structure has emerged \cite{qf1, qf2}.
The topological information computed are the Donaldson-Thomas invariants,
providing a powerful reformulation of Gromov-Witten invariants.
For topological string theories on Calabi-Yau manifolds there are additional
well-developed computational tools using open-closed duality such as the
topological vertex or matrix models.

In comparison, topological theories on $G_2$ manifold target
spaces are much less explored. One motivation to consider such
theories is since the $G_2$ structure couples K\"{a}hler and
complex structure naturally, such a theory would couple
topological A- and B-models non-perturbatively, a coupling which
we expected to exist following recent work on topological string
theory. A recent proposal for topological theories on $G_2$
manifolds that goes under the name of topological M-theory was
given in \cite{M}.

The classical effective description of topological M-theory is in
terms of a Hitchin functional \cite{hitchin1}. Alternative
topological theories on $G_2$ manifolds employing quantum
worldsheet/worldvolume formulations have been proposed in terms of
topological strings \cite{g2string} and topological membranes
\cite{membrane1, membrane2, membrane, membrane3}
\footnote{A topological version of F-theory on $Spin(7)$ manifolds
which are trivial torus fibrations over Calabi-Yau spaces was also
considered in \cite{topF}.}.
The topological $G_2$ string and topological membrane theories
\cite{membrane} have the same structure of local observables
associated to the de Rham cohomology of $G_2$ manifolds. The full
quantum worldvolume formulation of these theories, especially the
computation of the complete path integral is much more difficult
though than for the usual topological theories on Calabi-Yau
target spaces
\footnote{The topological $G_2$ string partition
function is only well-understood below genus two. At genus zero it computes the Hitchin
functional while its genus one contribution will be calculated in this paper. The topological membrane partition function is written only
formally.}
.

In this paper we attempt to understand the moduli space of
topological M-theory in terms of a $G_2$ target space description. Our strategy is similar to the A-model quantum foam, where one considers fluctuations around a fixed background K\"{a}hler form. Here the quantum path integral is computed in terms of a topologically twisted six-dimensional abelian gauge theory.

Analogously, the stable closed 3-form encoding the $G_2$ structure in seven dimensions can
be understood as a perturbation around a fixed background associative 3-form. Locally
the fluctuation can be regarded as the field strength of an abelian 2-form
gauge field. Unlike the A-model quantum foam, however, expanding the Hitchin functional
to quadratic order around this fixed background gives a seven-dimensional gauge theory that
is not quite topological but which is only invariant under diffeomorphisms of the $G_2$ manifold.

We will analyze the quantum structure of this theory by taking the 2-form gauge
field to be topologically trivial. In practise this means we will neglect certain \lq total derivative' terms in the
expansion of the Hitchin functional involving components of the bare 2-form gauge field
\footnote{For more conventional gauge theories such local \lq total derivative' terms usually correspond to topological invariants
computing certain characteristic classes for the gauge bundle from the patching conditions.
Unlike in conventional abelian gauge theory where the gauge field corresponds to a connection on a line bundle over the base space, the 2-form gauge field we have here corresponds to a connection on a gerbe.}
. This will allow us to generalize to seven dimensions the approach used by Pestun and Witten {\cite{PW}} to quantize the Hitchin functional for a stable 3-form in six dimensions to 1-loop order. This approach is based on the powerful techniques developed by Schwarz {\cite{SC}} for evaluating the partition function of a degenerate quadratic action functional. The structure of the partition function here is most naturally understood by fixing the gauge symmetry of the action using the antifield-BRST method of Batalin and Vilkovisky {\cite{BV}}. See {\cite{M,smo}} for possible alternatives to the perturbative quantization we consider here.

We will also investigate whether the 1-loop agreement found by Pestun-Witten {\cite{PW}} between the partition functions of the extended Hitchin functional in six dimensions and the topological B-model has some analogy in seven dimensions. In particular we will repeat the 1-loop partition function calculation for the extended Hitchin functional in seven dimensions to compare with the topological $G_2$ string. We find they are related only up to a multiplicative factor, corresponding to the Ray-Singer torsion invariant of the background $G_2$ manifold. It is not clear to us whether precise agreement could be obtained by a more careful analysis incorporating the global topological structure of the local total derivative terms we have dropped. Nonetheless, it seems that the topological symmetry of such terms could potentially give rise to non-trivial 1-loop determinants which we have ignored.

Our 1-loop quantization of the generalized $G_2$ Hitchin functional is in terms of linear variations of a closed stable odd-form in seven dimensions. However, the odd-form can be parameterized non-linearly in terms of other fields, that would be related to the dilaton, B-field, metric and RR flux moduli in compactifications of physical string theory on generalized $G_2$ manifolds. Hence an additional question is if we are using the appropriate degrees of freedom to describe the quantum theory. It would be interesting to see if our results could be checked by comparison with the couplings appearing in effective actions for generalized $G_2$ compactifications of physical string and M-theory.

A summary of the content of the paper is as follows. In section 2 we consider the expansion to cubic order of the Hitchin functional for a stable 3-form in seven dimensions around a fixed background $G_2$ manifold. It is only the quadratic term in the expansion that will contribute to the 1-loop partition function. The local total derivative terms will be identified in this quadratic action. Section 3 begins with a brief summary of the Lagrangian antifield-BRST or BV formalism followed by a detailed analysis of the BV quantization of the quadratic Hitchin action in seven dimensions. Section 4 begins by summarizing the theory of elliptic resolvents used in {\cite{SC}}. We will then identify the resolvent that describes the BV quantized seven-dimensional quadratic Hitchin action. This will allow us to express the partition function in terms of determinants of elliptic operators in the resolvent. In section 5 we will repeat the aforementioned analysis for the generalized Hitchin functional for a stable odd-form in seven dimensions. Section 6 contains a calculation of the 1-loop partition function of the topological $G_2$ string, which will be compared with the Hitchin functional computations. In section 7 these results will be compared with the Pestun-Witten analysis in one dimension lower. Section 8 describes the dependence of these 1-loop partition functions on the choice of background $G_2$ metric and proposes a seven-dimensional origin for the gravitational anomaly in the B-model. Section 9 contains our conclusions and a summary of interesting open questions related to this work.


\section{\large{Perturbation of Hitchin functional}} \label{sec-hf}
\setcounter{equation}{0}

Consider the proposed classical effective action for topological M-theory {\cite{M}} given by the Hitchin functional
\be
\frac{1}{7} \, \int_{M} \Phi \w *_{\Phi} \Phi \;,
\label{hf1}
\ee
for a stable closed 3-form $\Phi$, whose extrema within a given cohomology class of $\Phi$ define metrics of $G_2$ holonomy.
Recall that a Riemannian metric $g$ can be constructed from $\Phi$ using the formula
\be
\sqrt{g} \, g_{MN} \; =\; \frac{1}{144} \, \Phi_{MAB} \, \Phi_{NCD} \, \Phi_{EFG} \, \epsilon^{ABCDEFG} \; =: \; {\cal G}_{MN} \; ,
\ee
and it is with respect to this metric that the Hodge-star $*_\Phi$ is defined.
When both $\Phi$ and $*_{\Phi} \Phi$ are closed the Levi-Civita connection of the metric $g$ has holonomy in $G_2$ and ({\ref{hf1}})
corresponds to the volume
\be
{\mbox{vol}}_M \; =\; \int_M \sqrt{g} \; =\; \int_M ({\mbox{det}}\, {\cal G})^{1/9} \; ,
\ee
of the $G_2$ manifold $M$.

We will consider the expansion of the Hitchin functional ({\ref{hf1}}) around a fixed harmonic 3-form $\phi$ that encodes the geometry of
a background metric of $G_2$ holonomy, that is
\be
d \phi \; =\; 0 \; , \quad d {*\phi} \;=\; 0.
\ee
We define $M_0$ to be the seven-manifold $M$ equipped with this background metric.
For simplicity we take $\phi$ such that it reconstructs a flat metric although all subsequent formulae generalize in a
straightforward way for curved $G_2$ backgrounds.

Since $\Phi$ is closed then expanding around the fixed background
\be
\Phi \; =\; \phi + H \; ,
\ee
implies that $d H =0$. The 3-form perturbation $H$ is then understood as the
field strength of an abelian 2-form gauge
field $B$, which locally can be written as $H=dB$.

One can expand
\be
{\cal G}_{MN} \; =\; \delta_{MN} + {\cal A}_{MN} + {\cal B}_{MN} + {\cal C}_{MN} \; ,
\ee
in powers of $H$, such that
\bea
{\cal A}_{MN} &=& \frac{1}{4} \, H_{M}^{\;\;\; AB} \phi_{NAB} + \frac{1}{4} \, H_{N}^{\;\;\; AB} \phi_{MAB} \nonumber \\ [.1in]
{\cal B}_{MN} &=& \frac{1}{8} \, H_{MAB} \, H_{NCD} \, {*\phi}^{ABCD} \nonumber \\ [.1in]
{\cal C}_{MN} &=& \frac{1}{144} \, H_{MAB} \, H_{NCD} \, H_{EFG} \, \epsilon^{ABCDEFG} \label{abc} \; ,
\eea
where indices are contracted using the flat background metric.
To cubic order in $H$, one finds
\bea ({\mbox{det}}\, {\cal G})^{1/9} &=& 1+ \frac{1}{9} \, {\mbox{tr}} {\cal A} + \frac{1}{9} \left[ {\mbox{tr}} {\cal B} -\frac{1}{2} \, {\mbox{tr}}( {\cal A}^2 ) + \frac{1}{18} \, ({\mbox{tr}} {\cal A})^2 \right] \nonumber \\ [.1in]
&&+\frac{1}{9} \left[ {\mbox{tr}} \, {\cal C} - {\mbox{tr}}({\cal A}{\cal B}) + \frac{1}{9} \, {\mbox{tr}} {\cal A} \, {\mbox{tr}} {\cal B} \right. \nonumber \\ [.1in]
&&\hspace*{.4in}\left. +\frac{1}{3} \, {\mbox{tr}}( {\cal A}^3 ) -\frac{1}{18} \, {\mbox{tr}} {\cal A} \, {\mbox{tr}}( {\cal A}^2 ) + \frac{1}{486} \, ({\mbox{tr}} {\cal A})^3 \right] \label{hfex}  \; .
\eea
The unit term  corresponds to the volume measure for the flat background.
The linear term ${\mbox{tr}} {\cal A} = \frac{1}{2} \, \phi_{MNP} H^{MNP}$ is locally a total derivative which we will ignore in our analysis.
The quadratic and higher order terms become more complicated and so we will discuss their structure separately.


\subsection{Quadratic part}

Let us first note some useful identities that will allow us to write the quadratic terms in the Hitchin functional in a convenient way.
The projection operators in ({\ref{proj}}) can be used to decompose the 3-form
\bea
H_{MNP} &=& ( {\sf{P}}^3_{\bf{27}} H )_{MNP} + ( {\sf{P}}^3_{\bf{7}} H )_{MNP} + ( {\sf{P}}^3_{\bf{1}} H )_{MNP} \nonumber \\ [.1in]
&=& X_{MNP} + {*\phi}_{MNPQ} \, Y^Q + \phi_{MNP} \, Z \label{deco} \; ,
\eea
where $Y_M = \frac{1}{24} \, {*\phi}_{IJKM} H^{IJK}$ and $Z= \frac{1}{42} \, \phi_{IJK} H^{IJK}$.
The first identity in ({\ref{iden}}) then implies
\bea
2\, X_{MNP} &=& -3\, X^{IJ}_{\;\;\;\, [M} {*\phi}_{NP]IJ}  \\ [.1in]
(24)^2 \, Y_M Y^M + (42)^2 \, Z^2 &=& 6\, H_{MNP} H^{MNP} + 9\, {*\phi}^{IJKL} H_{IJM} H_{KL}^{\;\;\;\;\; M} \label{iden2} \nonumber  \; .
\eea
Substituting the decomposition ({\ref{deco}}) into the second term in the right hand side of ({\ref{iden2}}) gives
\be
H_{MNP} H^{MNP} \; =\; X_{MNP} X^{MNP} + 24\, Y_M Y^M + 42\, Z^2 \; .
\ee
This is useful because ${\cal A}_{MN} = \frac{1}{2} X_{M}^{\;\;\; AB} \phi_{NAB} + 3\, \delta_{MN} \, Z$ (using the identity $X_{M}^{\;\;\; AB} \phi_{NAB} = X_{N}^{\;\;\; AB} \phi_{MAB}$)
and explicit calculation gives
\be
{\mbox{tr}} ( {\cal A}^2 ) \; =\; \frac{1}{3} \, X_{MNP} X^{MNP} + 63\, Z^2 \; ,
\ee
so that using the previous identity implies
\bea
&&\hspace*{-.8in} {\mbox{tr}} {\cal B} -\frac{1}{2} \, {\mbox{tr}}( {\cal A}^2 ) + \frac{1}{18} \, ({\mbox{tr}} {\cal A})^2 \nonumber \\ [.1in]
&=&-\frac{1}{4} \, | H_{MNP} |^2 + \frac{1}{48} \, | {*\phi}_{MNPQ} H^{NPQ} |^2 + \frac{1}{72} \, | \phi_{MNP} H^{MNP} |^2 \label{quad1} \; .
\eea
One can check explicitly that the integral of this quadratic term is invariant under $\delta \phi_{MNP} = 0$, $\delta B_{MN} = \phi_{MNP} v^P$ (i.e. under background-preserving diffeomorphisms $\delta x^M = - v^M (x)$).

The quadratic term above is related to the metric on the moduli space of $G_2$ manifolds described by Hitchin in {\cite{NH}},
this metric being just the second functional derivative of ({\ref{hf1}}) with respect to $\Phi$.
In particular, if we define $\delta \Phi = H$ then ({\ref{quad1}}) can be used to write
\be
\frac{1}{9} \left[ {\mbox{tr}} {\cal B} -\frac{1}{2} \, {\mbox{tr}}( {\cal A}^2 ) + \frac{1}{18} \, ({\mbox{tr}} {\cal A})^2 \right] d^7 x \; =\; \frac{1}{6} \, \delta ( *_\Phi \Phi ) \wedge \delta \Phi \; ,
\label{quad2}
\ee
where
\be
\delta (*_\Phi \Phi ) \; =\; \frac{4}{3} *{\sf{P}}^3_{\bf{1}} ( \delta \Phi ) + *\, {\sf{P}}^3_{\bf{7}} ( \delta \Phi ) - *\, {\sf{P}}^3_{\bf{27}} ( \delta \Phi ) \; =:\; *{\sf L} ( \delta \Phi ) \; ,
\label{delphi}
\ee
to linear order in $\delta \Phi$, using the identities mentioned above and in appendix A
($*$ being the Hodge dual with respect to flat background $\phi$). The linear combination ${\sf L} = {4 \over 3} {\sf{P}}^3_{\bf{1}} + {\sf{P}}^3_{\bf{7}} - {\sf{P}}^3_{\bf{27}}$ of 3-form projectors has been defined for notational convenience in forthcoming expressions.
From this perspective it is clear that the $G_2$ moduli space metric has $(1+ b^3_{\bf 7} , b^3_{\bf 27} )$ signature.
This corresponds to the Lorentzian signature $(1, b^3 -1)$ for smooth compact seven-manifolds with full $G_2$ holonomy.

Like the linear term, some terms in the quadratic part of the Hitchin functional are also local total derivatives. To see this let us locally decompose the 2-form gauge field into irreducible representations of $G_2$
\be
B_{MN} \; =\; {\tilde B}_{MN} + {1 \over 6} \phi_{MNP} A^P \; ,
\ee
using the 2-form projection operators defined in ({\ref{proj}}), where ${\tilde B}_{MN} \in \Lambda^2_{\bf 14}$ and $A^M = \phi^{MNP} B_{NP}$. Plugging this expression into ({\ref{quad1}}) one finds
\bea
&&\hspace*{-.6in} {\mbox{tr}} {\cal B} -\frac{1}{2} \, {\mbox{tr}}( {\cal A}^2 ) + \frac{1}{18} \, ({\mbox{tr}} {\cal A})^2 \nonumber \\ [.1in]
&=&-\frac{1}{4} \, | {\tilde H}_{MNP} |^2 + \frac{1}{48} \, | {*\phi}_{MNPQ} {\tilde H}^{NPQ} |^2 \label{quad1a} \\ [.1in]
&&+ \partial_M \left[ {1 \over 6} \, A^{[M} \partial_N A^{N]} + {1 \over 24} \, {*\phi}^{MNPQ} A_N \partial_P A_Q - {1 \over 4} \, \phi^{MNP} {\tilde H}_{NPQ} A^Q  \right] \nonumber \; ,
\eea
where we define ${\tilde H} = d {\tilde B}$. This is just for notational convenience, ${\tilde H}$ is certainly not a gauge-invariant field strength. The integral of the second line in ({\ref{quad1a}}) is however gauge-invariant under $\delta {\tilde B} = {\sf P}^2_{\bf 14} d \lambda$. The third line in ({\ref{quad1a}}) involving the component $A$ in the ${\bf 7}$ irrep of $G_2$ is the aforementioned local total derivative. We will ignore these terms in the BV quantization in the next section. When $B$ is topologically trivial we are justified in neglecting them and their omission can be understood in terms of fixing diffeomorphism symmetry (by relating $A^M$ to the diffeomorphism generator $v^M$).

For the analysis in the next section it will be convenient to label the integral of the second line of ({\ref{quad1a}}) $S_0$ and rewrite it in form notation as
\be
S_0 \; =\; {3 \over 2} \int {\tilde H} \w * {\sf L} {\tilde H}  \; =\; {3 \over 2} \int {\tilde H} \w *\left( 2\, {\sf P}^3_{\bf 7} {\tilde H} - {\tilde H} \right) \; ,
\label{Scl}
\ee
which can be easily derived from ({\ref{quad2}}) and ({\ref{delphi}}) using the identity ${\sf P}^3_{\bf 1} {\tilde H} =0$.


\subsection{Cubic part}
The terms of cubic order in the Hitchin functional are most conveniently expressed in terms of the projections $X$, $Y$ and $Z$ of $H$ defined
in the previous subsection.
The new terms one must calculate (that are not just polynomials of the quadratic and linear ones) are
\bea
{\mbox{tr}} ( {\cal A}^3 ) &=& -{1\over 2} X_{MAB} X^{NAB} X^{MCD} \phi_{NCD} +  X_{MNA} X_{NPB} X_{PMC} \, \phi^{ABC} \nonumber \\
&& +3\, | X_{MNP} |^2 Z + 189\, Z^3 \nonumber \\ [.1in]
{\mbox{tr}} ( {\cal A} {\cal B} ) &=& -{3\over 8} X_{MAB} X^{NAB} X^{MCD} \phi_{NCD} + {1\over 2} \, X_{MNA} X_{NPB} X_{PMC} \, \phi^{ABC} \nonumber \\
&&+{2\over 3} \, | X_{MNP} |^2 Z - {1\over 2} X_{MAB} \phi^{NAB} Y^M Y_N + {1\over 7} \, {\mbox{tr}} ( {\cal A} ) \, {\mbox{tr}} ( {\cal B} ) \\ [.1in]
{\mbox{tr}} \, {\cal C}  &=& -{1\over 12} X_{MAB} X^{NAB} X^{MCD} \phi_{NCD} + {1\over 6} \, X_{MNA} X_{NPB} X_{PMC} \, \phi^{ABC} \nonumber \\
&&-{1\over 12} \, | X_{MNP} |^2 Z + X_{MAB} \phi^{NAB} Y^M Y_N + 6\, | Y_M |^2 Z + 7\, Z^3 \nonumber \; .
\eea

Combining these expressions one finds
\bea
&&{\mbox{tr}} \, {\cal C} - {\mbox{tr}}({\cal A}{\cal B}) + \frac{1}{9} \, {\mbox{tr}} {\cal A} \, {\mbox{tr}} {\cal B} +\frac{1}{3} \, {\mbox{tr}}( {\cal A}^3 ) -\frac{1}{18} \, {\mbox{tr}} {\cal A} \, {\mbox{tr}}( {\cal A}^2 ) + \frac{1}{486} \, ({\mbox{tr}} {\cal A})^3  \nonumber \\ [.1in]
&&\hspace*{.2in} = {1\over 8} X_{MAB} X^{NAB} X^{MCD} \phi_{NCD} + {3\over 2} \, X_{MAB} \phi^{NAB} Y^M Y_N  \nonumber \\ [.1in]
&&\hspace*{.3in} - {1\over 12} \, | X_{MNP} |^2 Z + 2\, | Y_M |^2 Z + {14\over 9} \, Z^3 \; .
\eea

This diffeomorphism invariant cubic term can be understood as a
BRST invariant operator deforming the quadratic action calculated
above. It will not effect the 1-loop calculation of the partition
function we are interested in here but is important when going
beyond this order.


\section{\large{BV quantization}} \label{sec-bv}
\setcounter{equation}{0}

Before getting into the details of the BV quantization of the quadratic Hitchin action, it may be helpful to set up terminology by first giving a brief review of the Lagrangian antifield-BRST formalism, following the excellent lectures by Henneaux {\cite{Hen}} where more details can be found.


\subsection{Lagrangian antifield-BRST formalism}

The basic idea is to implement the restriction of the configuration space spanned by functions of all fields $\{ \phi \} \in I$ in a given field theory with classical action $S_0 [ \phi ]$ to the physical subspace of functions of on-shell configurations $\Sigma \subset I$ modulo gauge-equivalence, by means of constraints involving a nilpotent BRST operator $Q$ on the former space giving the latter space as its cohomology. The construction of $Q$ can be understood in terms of two preliminary nilpotent operators $\delta$ and $d$
\footnote{This notation will be used exclusively in this subsection and is not to be confused with general infinitesimal variations labeled $\delta$ and spacetime exterior derives labeled $d$ elsewhere in the paper.}
which are used to individually impose the on-shell and gauge-equivalence constraints respectively.

The zeroth homology group $H_0 ( \delta )$ is known as the {\emph{Koszul-Tate resolution}} of $C^\infty ( \Sigma )$. It turns out to be sufficient to consider $H_0 ( \delta )$ due to a consistency condition implying that all higher homology groups $H_{k>0} ( \delta )$ must vanish. The grading of $\delta$ is called {\emph{antighost number}} ($\{ \phi \}$ have antighost number zero while $\delta$ itself has antighost number $-1$ which is why the on-shell configuration space is a homology rather than cohomology group). In the absence of gauge symmetry, the structure of $H_0 ( \delta )$ is necessarily simple. The kernel $({\mbox{ker}}\, \delta )_0 = C^\infty (I)$. The image $({\mbox{im}}\, \delta )_0$ is the subset of functions whose vanishing defines all the equations of motion. Thus one has implicitly defined a set of so-called {\emph{antifields}} $\{ \phi^* \}$ such that $\delta$ acting on each antifield $\phi^*$ is proportional to the field equation for $\phi$ (thus $\{ \phi^* \}$ have antighost number 1).

In the presence of gauge symmetry one still has $H_0 ( \delta ) = C^\infty ( \Sigma )$ but now one finds $H_{k>0} (\delta ) \neq 0$ which contradicts the consistency condition. The reason for this is because gauge transformations of the antifields $\{ \phi^* \}$ are closed under $\delta$ and thus lie in $H_1 (\delta )$. The trick is to introduce more so-called {\emph{antighost fields}} $\{ C^* \}$ (with antighost number 2) such that each of the $\delta$-closed irreducible gauge variations above equals $\delta C^*$ and is therefore trivial in $H_1 (\delta )$. This turns out to be sufficient if all gauge symmetries are irreducible (i.e. if no possible gauge transformations of fields vanish or are proportional to equations of motion). For each reducible gauge symmetry one can have a non-trivial element of $H_{2} (\delta )$ which is again avoided by adding another antighost field $\eta^*$ (with antighost number $3$) to trivialize it. Obstructions to the triviality of $H_{k>2} (\delta )$ will not concern us here and we refer the interested reader to {\cite{Hen}} for a discussion of their resolution.

The zeroth cohomology group $H^0 (d)$ corresponds to the algebra of gauge-invariant functions on $\Sigma$. Non-vanishing higher cohomology groups $H^{k>0} (d) \neq 0$ are allowed. The grading of $d$ is called {\emph{pure ghost number}} ($\{ \phi \}$ have pure ghost number zero while $d$ itself has pure ghost number $1$). The action of the vertical exterior derivative along gauge orbits $d$ is generated by the set of tangent vectors $\{ X \}$ at $\{ \phi \}$ on $\Sigma$. The number of such tangent vectors equals the number of linearly independent gauge symmetries. It is convenient to define the set of 1-forms or {\emph{ghosts}} $\{ C \}$ as the dual of the tangent vectors $\{ X \}$ (thus $\{ C \}$ have pure ghost number 1). In the case where there exist reducible gauge symmetries, the set $\{ X \}$ form an overcomplete basis since its elements are subject to linear algebraic constraints on $\Sigma$ (one constraint per reducible symmetry). It turns out one can enforce these constraints automatically by modifying the action of $d$ on $\{ C \}$ in terms of additional {\emph{ghost for ghost}} fields $\eta$ (one per reducible symmetry with pure ghost number 2) to create a free differential algebra.

Collecting all these fields together we see that there is a perfect match between the number of (fields+ghosts) $\Phi = \{ \phi , C , \eta \}$ and anti(fields+ghosts) $\Phi^* = \{ \phi^* , C^* , \eta^* \}$. The physical BRST operator $Q$ acts on both $\Phi$ and $\Phi^*$ and its cohomology can be understood as the cohomology of $d$ on $\Sigma$. Schematically one has $Q = d+\delta + {\mbox{\lq extra'}}$ and it turns out one can always choose \lq extra' such that $Q^2 =0$. For relatively simple abelian gauge theories like the one we consider there are no \lq extra' terms. The grading of $Q$ is called {\emph{ghost number}} which, from the formula above, is given by the pure ghost number minus the antighost number. Hence $\{ \eta^* , C^* , \phi^* , \phi , C, \eta \}$ have ghost numbers $\{ -3,-2,-1,0,1,2 \}$. Since $Q$ is a fermionic nilpotent operator then $\{ \phi, C^* , \eta \}$ obey bosonic statistics while $\{ \phi^* , C, \eta^* \}$ obey fermionic ones (we have assumed the original fields $\{ \phi \}$ are all bosonic).

The pairing between $\Phi$ and $\Phi^*$ implies the existence of a graded symplectic structure on the space of fields given by the {\emph{antibracket}}
\footnote{It must be stressed that the antibracket in Lagrangian formalism is not induced from the Poisson bracket in Hamiltonian formalism. The antibracket seems to be a purely auxiliary structure that is lost when one fixes a gauge.}
\be
(A,B) \; =\; {\delta^r A \over \delta \Phi} \cdot {\delta^l B \over \delta \Phi^*} - {\delta^r A \over \delta \Phi^*} \cdot {\delta^l B \over \delta \Phi} \; ,
\ee
where $A$ and $B$ are arbitrary functionals of both $\Phi$ and $\Phi^*$. The symbol $\cdot$ denotes summation over all common indices of all fields in $\Phi$ and $\Phi^*$. For the theories we will consider, elements of $\Phi$ will be in form representations and it will sometimes be more convenient henceforth to take elements in $\Phi^*$ to be in the Hodge-dual representations to their partners in $\Phi$. Thus we would understand the antibracket above as the coefficient of a top-form in spacetime and replace the contraction of common indices $\cdot$ with a wedge product . The superscripts on the functional derivatives denote a right ($r$) and left ($l$) action on $A$ and $B$, which is required by the grading.

The antibracket is useful because it allows the construction of the minimal BRST-invariant action $S[ \Phi , \Phi^* ]$, involving ghosts and antifields, that includes $S_0 [ \phi ]$. This is achieved by solving the {\emph{master equation}}
\be
Q \, F \; =\; (F, S ) \; ,
\label{master}
\ee
for any functional $F$. Nilpotence of $Q$ ensures that $( S , S ) =0$. An immediate consequence of the master equation is that $Q \Phi = {\delta^l S \over \delta \Phi^*}$ and $Q \Phi^* = -{\delta^l S \over \delta \Phi}$. The proper solution $S$ is referred to as minimal because one can always introduce new variables $\{ {\bar C} , \pi \} $ (and their respective antifields $\{ {\bar C}^* , \pi^* \}$) that are cohomologically trivial in $H^0 (Q)$ (i.e. $Q {\bar C} = \pi$ , $Q \pi =0$, $Q \pi^* = {\bar C}^*$, $Q {\bar C}^* = 0$) and which do not contribute to $H^{k>0} (Q)$. Thus one can add terms of the form $\int {\bar C}^* \cdot \pi$ to $S$ to obtain the most general solution of the master equation. Such non-minimal terms typically arise in the process of gauge-fixing where the antifields $\{ {\bar C}^* \}$ are related to the gauge-fixing functions and $\{ \pi \}$ act as Lagrange multipliers imposing ${\bar C}^* =0$.

The final step is to remove the degeneracy (i.e. gauge symmetry) in the action above in a way that preserves the BRST structure, which will allow a more straightforward evaluation of the path integral. If there are $2N$ fields in $\{ \Phi , \Phi^* \}$ then this gauge-fixing can be achieved by eliminating half of them via $N$ constraints $\{ \Omega = 0 \}$. Such constraints are guaranteed to preserve the BRST structure provided the antibracket of any two $\Omega$ in the set vanishes
\footnote{That is their antibracket is invariant under canonical graded symplectic transformations which define the ambiguity in determining the minimal action $S$ from the master equation.}
. A convenient way to satisfy the above constraint is to eliminate all the antifields by setting each $\Phi^* = {\delta^r \Psi \over \delta \Phi}$ for some choice of {\emph{gauge fermion}} functional $\Psi [ \Phi ]$ (this choice is by no means unique). It is evident from the definition that $\Psi$ must be fermionic and have ghost number $-1$. Notice that this constraint has removed the antibracket structure for the gauge-fixed theory. The aforementioned gauge choice can be understood geometrically as restricting to a Lagrangian submanifold of the symplectic manifold parameterized by $\{ \Phi , \Phi^* \}$ (these terms of course being used in the graded sense).


\subsection{BV quantization of quadratic Hitchin action}

We are now prepared to examine the quantum structure of ({\ref{Scl}}) for ${\tilde B} \in \Lambda^2_{\bf 14}$ following the logic of the previous subsection.

In addition to the 2-form gauge field ${\tilde B}$ we need a 1-form fermionic ghost $\psi$ and a 0-form bosonic ghost-for-ghost $\varphi$. The ghost comes from the 1-form $\lambda$ which parameterizes the gauge symmetry $\delta {\tilde B} = {\sf P}^2_{\bf 14} d \lambda$ of $S_0 [ {\tilde B} ]$. This gauge symmetry is reducible when $\lambda = d \kappa$ for any 0-form $\kappa$, giving rise to ghost-for-ghost $\varphi$. The antifields for $\Phi = \{ {\tilde B} , \psi , \varphi \}$ are $\Phi^* = \{ {\tilde \chi} , \zeta , \xi \}$ which lie in the Hodge-dual irreps of $G_2$. That is ${\tilde \chi}$ is a 5-form fermion whose Hodge-dual is in $\Lambda^2_{\bf 14}$, $\zeta$ is a 6-form boson and $\xi$ is a 7-form fermion. The ghost numbers of $\{ \xi , \zeta , {\tilde \chi} , {\tilde B} , \psi , \varphi \}$ are $\{ -3,-2,-1,0,1,2 \}$ respectively.

The global BRST transformations of the fields and ghosts $\Phi$ just follow from the residual local gauge transformations
\be
Q {\tilde B} \; =\; {\sf P}^2_{\bf 14} d \psi \; , \quad Q \psi \; =\; d\varphi \; , \quad Q \varphi \; =\; 0 \; .
\ee
The master equation $Q \Phi = {\delta S \over \delta \Phi^*}$ then fixes the terms one must add to the classical action $S_0$ ({\ref{Scl}}) to be of the form $\int \Phi^* \w Q \Phi$. Thus the minimal solution to the master equation ({\ref{master}}) is
\be
S \; =\; \int {3 \over 2} \, {\tilde H} \w *\left( 2\, {\sf P}^3_{\bf 7} -1 \right) {\tilde H} + {\tilde \chi} \w d \psi + \zeta \w d \varphi \; .
\label{Smin}
\ee
The projector ${\sf P}^2_{\bf 14}$ in $Q {\tilde B}$ has been absorbed by ${*{\tilde \chi}} \in \Lambda^2_{\bf 14}$ in the second term in ({\ref{Smin}}). Using ({\ref{Smin}}) in the other master equation $Q \Phi^* = -{\delta S \over \delta \Phi}$ gives the antifield BRST transformations
\be
Q {\tilde \chi} \; =\; 3\, {d* \left( 2 {\sf P}^3_{\bf 7} -1 \right)} d {\tilde B} \; , \quad Q \zeta \; =\; d {\tilde \chi} \; .
\ee
One can verify that the above BRST transformations indeed generate a symmetry of $S$ and obey $Q^2 =0$. One can also check that the BRST transformation $Q {*{\tilde \chi}}$ is in $\Lambda^2_{\bf 14}$ as required. Notice that any BRST transformation of $\xi$ will be a symmetry of $S$, and will be nilpotent provided it is BRST-trivial (i.e. $Q \xi = \mu$, $Q \mu =0$).

To fix the gauge symmetry of the field ${\tilde B}$ and ghost $\psi$ via the constraints $d^\dagger {\tilde B} =0$ and $d^\dagger \psi =0$ it is appropriate to add to $S$ some non-minimal terms via the introduction of the pair of 6-forms $\{ \gamma , u \}$ and 7-forms $\{ \varepsilon , v \}$ (plus their antifield 1-forms $\{ \gamma^* , u^* \}$ and 0-forms $\{ \varepsilon^* , v^* \}$) which are BRST-trivial (i.e. $Q \gamma = u$, $Q u =0$ etc.). The appropriate gauge fermion in this case is given by
\be
\Psi \; =\; \int \gamma \w d^\dagger {\tilde B} + \varepsilon \w d^\dagger \psi + \gamma \w d \theta \; .
\ee
The first two terms are as we would expect in order to gauge fix via coclosure of ${\tilde B}$ and $\psi$. The reason for the third term involving an additional BRST-trivial 0-form pair $\{ \theta , w \}$ is because it fixes a residual gauge symmetry of the first term under $\delta \gamma = d^\dagger \rho$ for any fermionic 7-form $\rho$. There is no possible residual symmetry from the second term since $\delta \varepsilon$ cannot be coexact in seven dimensions. The corresponding BRST-invariant non-minimal addition to the action $S$ is
\be
\int \gamma^* \w u + \varepsilon^* \w v + \theta^* \w w \; .
\ee
Thus $\{ u,v,w \}$ are understood as auxiliary fields (with ghost numbers \\ $\{ 0,-1,1 \}$) that will impose the gauge-fixing constraints after imposing $\Phi^* = {\delta \Psi \over \delta \Phi}$ on the antifields.

Including the non-minimal fields we have $\Phi = \{ {\tilde B} , \psi , \varphi ; \gamma , \varepsilon , \theta \}$ (with ghost numbers $\{ 0,1,2;-1,-2,0 \}$) and $\Phi^* = {\delta \Psi \over \delta \Phi}$ fixes the antifields to be
\bea
{\tilde \chi} &=& *{\sf P}^2_{\bf 14} {*d^\dagger} \gamma \; , \quad \zeta \; =\; d^\dagger \varepsilon \; , \quad \xi \; =\; 0 \; ; \nonumber \\ [.1in]
\gamma^* &=& d^\dagger {\tilde B} + d \theta \; , \quad \varepsilon^* = d^\dagger \psi \; , \quad \theta^* \; =\; d \gamma \; .
\label{gfix}
\eea
The antifields of the Lagrange multipliers $\{ u,v,w \}$ all vanish. Integrating out these auxiliary fields in the non-minimal part of the action sets the three expressions in the second line of ({\ref{gfix}}) equal to zero. This evidently gives the desired gauge-fixing for $\psi$. Taking $d^\dagger$ of $d^\dagger {\tilde B} + d \theta =0$ implies $\theta$ must be harmonic and thus equal to a constant (we assume $\theta$ is non-singular). Hence we also have $d^\dagger {\tilde B} =0$.

In the topologically trivial case we are considering, these equations further imply the global constraints that $\gamma$ be exact and $\psi$ be coexact. This of course follows from the Poincar\'{e} lemma which for ${\tilde B} \in \Lambda^2_{\bf 14}$ is a bit more subtle. Indeed $d^\dagger {\tilde B} =0$ still implies ${\tilde B} = d^\dagger \Xi$, for some 3-form $\Xi$, but now there is the additional constraint ${\sf P}^2_{\bf 7} d^\dagger \Xi =0$ so that the right hand side is still in $\Lambda^2_{\bf 14}$. This is non-trivial because exterior derivatives do not commute with projection operators. As shown in appendix C, this leads to an expression for ${\tilde B}$ that is second order in derivatives
\footnote{An analogous situation occurs in K\"{a}hler geometry where the existence of a coclosed real (1,1)-form $b_{11}$ (obeying $\partial^\dagger b_{11} = 0$ and ${\bar \partial}^\dagger b_{11} =0$) implies $b_{11} = \partial^\dagger {\bar \partial}^\dagger \alpha_{22}$, for some real (2,2)-form $\alpha_{22}$. This example is used by Pestun and Witten {\cite{PW}} in gauge-fixing the quadratic Hitchin functional in six dimensions.}
. Let us then summarize the gauge-fixing constraints (just quoting the result shown in appendix C)
\bea
{\tilde B} &=& d^\dagger \left( 2\, {\sf P}^3_{\bf 7} -1 \right) d {\tilde \alpha} \; , \quad \psi \; =\; d^\dagger \beta \; , \nonumber \\ [.1in]
{\tilde \chi} &=& *{\sf P}^2_{\bf 14} d d^\dagger \omega \; , \quad \zeta \; =\; d^\dagger \varepsilon \; , \quad \xi \; =\; 0 \; ,
\label{gfix2}
\eea
where ${\tilde \alpha} \in \Lambda^2_{\bf 14}$, $\beta, \omega \in \Lambda^2$ and $\varepsilon \in \Lambda^7$. Plugging these expressions into the graded symplectic form
\be
\int \Phi^* \w \Phi \; =\; \int {\tilde \chi} \w {\tilde B} + \zeta \w \psi + \xi \w \varphi \; ,
\ee
for the minimal fields implies it vanishes identically, thus defining a Lagrangian submanifold. From this we see how the strong constraint $\xi =0$ is required by the fact that $\varphi$ is completely unconstrained.


\section{\large{Resolvent and partition function}} \label{sec-respar}
\setcounter{equation}{0}

Having gauge-fixed the quadratic part of the Hitchin action in a way that preserves the BRST structure, we are now almost prepared to evaluate its partition function. We will express the partition function in terms of determinants of elliptic operators using the theory of resolvents developed by Schwarz {\cite{SC}}. Again to introduce the necessary terminology it will be helpful to give a very brief account of the theory of resolvents as described in {\cite{SC}} wherein we defer for more a detailed exposition.


\subsection{Resolvents}

A resolvent is a generalization of a complex in algebraic geometry that is associated with an additional piece of data corresponding to a quadratic functional on one of the linear vector spaces in the complex. This functional will be understood as the classical action for a free field theory.

More precisely, given a quadratic functional $S_0$ on a linear space $\Gamma_0$, then a sequence of linear spaces $\Gamma_i$ ($i=1,...,n$) and nilpotent linear operators $T_i : \Gamma_i \rightarrow \Gamma_{i-1}$ obeying $T_{i-1} T_i =0$ is defined to be the {\emph{resolvent}} of $S_0$ if $S_0 [ \phi + T_1 C ] = S_0 [ \phi ]$ for all $\phi \in \Gamma_0$ and $C \in \Gamma_1$. This defines a complex when $S_0 =0$. The notation reflects that used in section 3.1 to illustrate that $\Gamma_i$ are to be understood as being spanned by all the descendent ghosts for classical bosonic fields $\phi$ in the quantum theory. The resolvent property corresponds to BRST symmetry.

The existence of an inner product $\langle \; ,\, \rangle$ on the linear spaces $\Gamma_0$ and $\Gamma_i$ will be assumed and adjoint linear operators $T^\dagger_i : \Gamma_i \rightarrow \Gamma_{i+1}$ can be constructed from $\langle x , T_i \, y \rangle = \langle T^\dagger_i x , y \rangle$ for any $x \in \Gamma_{i-1}$, $y \in \Gamma_i$. It will also be assumed that the quadratic functional $S_0$ can be expressed schematically as
\be
S_0 [ \phi ] \; =\; \langle \phi , K \phi \rangle \; =\; \langle K \phi , \phi \rangle \; ,
\ee
in terms of the self-adjoint \lq kinetic' operator $K : \Gamma_0 \rightarrow \Gamma_0$
\footnote{A further technical requirement is that the operators $K^2$ and $T^\dagger_i T_i$ be {\emph{regular}}.
We refer to {\cite{SC}} for the technical definition of regularity but the upshot is that this allows one to define the
(regularized) determinant of such operators in a mathematically precise way.}
. Note that the resolvent property implies $K T_1 =0$ and so $K$ itself can be added to the complex associated to the resolvent as
\be
0 \longrightarrow \Gamma_n \overset{T_n}{\longrightarrow} .\, .\, .\, \overset{T_1}{\longrightarrow} \Gamma_0 \overset{K}{\longrightarrow} \Gamma_0 \longrightarrow 0 \; .
\ee
The resolvent of $S_0$ is said to be {\emph{elliptic}} if the associated complex above is elliptic (i.e. the symbols of each $T_i$ and $K$ are invertible).

The {\emph{partition function}} of $S_0$ with respect to the resolvent $\{ \Gamma_i , T_i \}$ is defined
\be
Z \; =\; ({\mbox{det}}\, K )^{-1/2} \prod_{i=1}^{n} |{\mbox{det}}\, T_i |^{(-1)^{i-1}} \; .
\label{partf}
\ee
The reason that this quantity corresponds to the physical partition function for theory with classical action $S_0$ is explained
in the appendix of the third reference in {\cite{SC}}.
The $({\mbox{det}}\, K )^{-1/2}$ factor of course just comes from the path integral of the free bosonic action $S_0$.
In terms of the antifield-BRST formalism, the remaining ghost and antifield terms in the minimal action $S$ solving the master equation
take the form $\langle \phi^* , T_1 C \rangle + \langle C^* , T_2 \eta \rangle +...$. That is schematically
$\sum_{i=1}^n \langle \Gamma^*_{i-1} , T_i \Gamma_i \rangle$, where elements of $\Gamma_i$ have Grassmann parity $(-1)^{i}$ and
$\Gamma^*_i$ is the same vector space as $\Gamma_i$ but with elements of opposite Grassmann parity.
This leads to a factor $({\mbox{det}}\, T_i )^{(-1)^{i-1}/2}$ from the antifields in each $\Gamma^*_{i-1}$
and a factor $({\mbox{det}}\, T^\dagger_i )^{(-1)^{i-1}/2}$ from the ghosts in each  $\Gamma_i$ which are combined to give ({\ref{partf}}).


\subsection{$G_2$ resolvent for quadratic Hitchin action}

Given the close relationship between complexes and resolvents one might expect the resolvent for the quadratic Hitchin action we are
considering to be related to the two well-known Dolbeault complexes for $G_2$ manifolds
\bea {\check D} &:& 0 \longrightarrow \Lambda^0_{\bf 1}
\overset{d}{\longrightarrow} \Lambda^1_{\bf 7} \overset{{\sf
P}^2_{\bf 7} d}{\longrightarrow} \Lambda^2_{\bf 7} \overset{{\sf
P}^3_{\bf 1} d}{\longrightarrow} \Lambda^3_{\bf 1} \longrightarrow
0 \nonumber \\ [.1in] {\tilde D} &:& 0 \longrightarrow
\Lambda^2_{\bf 14} \overset{d}{\longrightarrow} \Lambda^3_{\bf 7}
\oplus \Lambda^3_{\bf 27} \overset{{\sf P}^4_{{\bf 7} \oplus {\bf
27}}d}{\longrightarrow} \Lambda^4_{\bf 7} \oplus \Lambda^4_{\bf
27} \overset{{\sf P}^5_{\bf 14} d}{\longrightarrow} \Lambda^5_{\bf
14} \longrightarrow 0 \; . \label{g2dol} \eea
As we will now see, the appropriate complex for the $G_2$ Hitchin action forms a subset of both these complexes.

Following the discussion in the previous subsection for the quadratic Hitchin action ({\ref{Scl}}) we identify $\Gamma_0 = \Lambda^2_{\bf 14}$.
For suitable normalization of ${\tilde B}$, the kinetic operator in $S_0$ is
\be
K \; =\; - d^\dagger {\sf L} d \; =\; \Delta^2_{\bf 14} - {3\over 2}\, {\sf P}^2_{\bf 14} d d^\dagger \; ,
\ee
which is self-adjoint and indeed maps $\Lambda^2_{\bf 14} \rightarrow \Lambda^2_{\bf 14}$. The later fact follows from the right hand side of the expression above or by noting the identity ${\sf P}^2_{\bf 7} d^\dagger {\sf L} d {\sf P}^2_{\bf 14} =0$
\footnote{For any $\beta \in \Lambda^2_{\bf 14}$, one can use the identity ${\sf P}^3_{\bf 7} d \beta = -{1 \over 4} *( \phi \w d^\dagger \beta )$ to derive this result. It is obtained by first noting that ${\sf L} d \beta = ( 2 {\sf P}^3_{\bf 7} -1 ) d \beta$, then substituting $2 d^\dagger {\sf P}^3_{\bf 7} d \beta = -{1 \over 2} *( \phi \w d d^\dagger \beta ) = - d d^\dagger \beta + {3 \over 2} {\sf P}^2_{\bf 14} d d^\dagger \beta$ that follows from taking $d^\dagger$ of the aforementioned identity. The second equality here follows from the 2-form projector identities $* {\sf P}^2_{\bf 7} = {1 \over 2} \phi \w {\sf P}^2_{\bf 7}$ $* {\sf P}^2_{\bf 14} = - \phi \w {\sf P}^2_{\bf 14}$, which can be derived from the expressions in appendix B.}
. The structure of ghosts encountered in section 3.2 implies $n=2$ with $T_1 = {\sf P}^2_{\bf 14}\, d$ and $T_2 =d$
(their adjoints being just $T^\dagger_1 = d^\dagger$ and $T^\dagger_2 = d^\dagger$).
Hence the appropriate complex is
\be
{\hat D} \;\; :\;\; 0 \longrightarrow \Lambda^0_{\bf 1} \overset{d}{\longrightarrow} \Lambda^1_{\bf 7} \overset{{\sf P}^2_{\bf 14} d}{\longrightarrow} \Lambda^2_{\bf 14} \overset{K}{\longrightarrow} \Lambda^2_{\bf 14} \longrightarrow 0 \; ,
\label{g2res}
\ee
with the extension by $K$ included.
The resolvent properties can be checked explicitly but of course just follow from the BRST structure. Notice that the first two elements match those in the ${\check D}$ complex in ({\ref{g2dol}}) while the third element corresponds to the first element in the ${\tilde D}$ complex.

Using these identifications, the partition function
({\ref{partf}}) for ({\ref{Scl}}) can be written
\bea Z &=& \left({\mbox{det}}\, \left( \Delta^2_{\bf 14} - {3 \over 2} {\sf P}^2_{\bf 14} d_1 d^\dagger_2 {\sf P}^2_{\bf 14} \right) \right)^{-1/2} |{\mbox{det}}\, ( {\sf P}^2_{\bf 14} d_1 ) ||{\mbox{det}}\, ( d_0 ) |^{-1} \nonumber \\ [.1in]
&=& ({\mbox{det}}\, \Delta^2_{\bf 14} )^{-1/2} |{\mbox{det}}\, ( {\sf P}^2_{\bf 14} d_1 ) |^{2} ({\mbox{det}}\, \Delta^0_{\bf 1} )^{-1/2} \nonumber \\ [.1in]
&=& ({\mbox{det}}\, \Delta^2_{\bf 14} )^{-1/2} ( {\mbox{det}}\, \Delta^1_{\bf 7} ) ({\mbox{det}}\, \Delta^0_{\bf 1} )^{-3/2} \; .
\label{partf2}
\eea
Superscripts (subscripts) denote the form degree ($G_2$ irrep) on which Laplacian operator $\Delta^i = d^\dagger_{i+1} d_i + d_{i-1} d^\dagger_i$ acts.
This action is invariant because $\Delta$ commutes with the projection operators on any $G_2$ manifold
\footnote{This is not entirely obvious but can be deduced from the fact that $G_2$ manifolds are Ricci-flat $R_{MN} =0$ and their
Riemann tensor obeys $R_{MNPQ} \phi^{PQA} =0$.
The latter property can be deduced from the formula $[ \nabla_M , \nabla_N ] \xi = {1 \over 4} R_{MNAB} \Gamma^{AB} \xi$ for covariant
derivatives $\nabla_M$ acting on the spinor $\xi$. That is the gamma matrices $\Gamma_{AB}$ on the right hand side must generate the
$G_2 \subset SO(7)$ holonomy group of the manifold and so only those $[AB]$ indices in the adjoint ${\bf 14}$ of $G_2$ should appear
on the right hand side of the commutator. Thus the ${\bf 7}$ part of the $[AB]$ indices of $R_{MNAB}$ must vanish identically which gives
the desired property.}
. The second equality has been obtained using the identity
${\mbox{det}}\, ( K + T_1 T_1 ^\dagger ) = ( {\mbox{det}}\, K) |
{\mbox{det}}\, T_1 |^{2}$ which follows because $K T_1 T^\dagger_1
=0$ and $T_1 T^\dagger_1 K = 0$. The final equality is obtained
using a similar result ${\mbox{det}}\, ( T^\dagger_1 T_1 + T_2
T^\dagger_2 )= | {\mbox{det}}\, T_1 |^2 | {\mbox{det}}\, T_2 |^2$
following from $T^\dagger_1 T_1 T_2 T^\dagger_2 = 0$ and $T_2
T^\dagger_2 T^\dagger_1 T_1 =0$.

We have omitted the infinite-volume normalization factors coming
from the zero modes of the Laplacians above. Formally the
multiplicative factor from these zero modes can be written
${\mbox{Vol}} ( H^2_{\bf 14} ){\mbox{Vol}} ( H^0_{\bf 1} ) /
{\mbox{Vol}} ( H^1_{\bf 7} )$ in terms of \lq volumes' of the
appropriate cohomology groups.


\section{Generalized Hitchin functional} \label{gen-hitchin}
\setcounter{equation}{0}

In \cite{PW}, one-loop computations in a theory based on the
six-dimensional Hitchin functional were compared with one-loop
computations in topological string theory, and the results were
found to disagree. However, agreement was found once the Hitchin
functional, which is a functional of a stable 3-form, was replaced
by the generalized Hitchin functional \cite{NHN}, which is a
functional of a generic stable form of odd degree. We will now
repeat the analysis of the previous sections for an analogous
generalization of the Hitchin functional in seven dimensions. The
result of \cite{PW} suggests it is this theory that should be
compared with the topological $G_2$ string.

\subsection{The generalized Hitchin functional}

The appropriate generalization of the Hitchin functional in seven
dimensions was described and studied in \cite{wi1,wi2,wi3}. Its
critical points correspond to seven-manifolds $M$ with
{\emph{generalized $G_2$ structure}}. That is, the structure group
$Spin(7,7)$ of $TM \oplus T^* M$ is reduced to $G_2\times G_2$.
This $G_2 \times G_2$ is the stabilizer of a generic form of odd
degree in seven dimensions under the action of the conformal
structure group $Spin(7,7) \times {\mathbb{R}}^*$. Each $G_2
\subset Spin(7)$ fixes a unit spinor on $M$. For a fixed embedding
$Spin(7) \times Spin(7) \subset Spin(7,7) \times {\mathbb{R}}^*$,
one finds the generalized $G_2$ structure reduces to an ordinary
one when these two spinors are parallel. We defer to
\cite{wi1,wi2,wi3} for a more detailed discussion.

The explicit construction of the generalized Hitchin functional
proceeds as follows. One begins by writing a stable odd-form
$\varrho\in \Lambda^{\rm odd} \cong \Lambda^1  \oplus
\Lambda^3  \oplus \Lambda^5 \oplus \Lambda^7$
\footnote{Stability here means that the orbit of $\varrho$ under
$Spin(7,7) \times {\mathbb{R}}^*$ forms an open subset of
$\Lambda^{\rm odd}$.}
as
\footnote{We define $e^{\sf B} = 1 + {\sf B} + {1 \over 2} {\sf
B}\w {\sf B} + {1 \over 6} {\sf B}\w {\sf B}\w {\sf B}$ in seven
dimensions.}
\be \label{defrho} \varrho \; =\; e^{-\varphi} e^{\sf B} \wedge
\left( s \alpha -c\, \Phi -s {*_\Phi (\alpha\wedge \Phi)} - s
\alpha\wedge {*_\Phi \Phi} + c\, \frac{1}{7} \Phi \wedge {*_\Phi
\Phi} \right) \; , \ee
in terms of a scalar \lq dilaton' $\varphi$, a 2-form ${\sf B}$
(which is related to the $B$-field) and a 3-form $\Phi$. The Hodge
star $*_\Phi$ is defined in terms of the metric associated with
$\Phi$, as in section {\ref{sec-hf}}. Furthermore, $s$ and $c$ are
real numbers satisfying $s^2 + c^2 =1$ and $\alpha$ is a unit
1-form, i.e. $\int_M \alpha\wedge *_\Phi \alpha= \int_M
\frac{1}{7} \Phi \wedge *_\Phi \Phi$. Despite the highly
non-linear structure in (\ref{defrho}), a simple
consistency check verifies that the number of independent
components on the left- and right-hand side match. The odd-form
$\varrho$ has $\sum_{p=0}^3 {7 \choose 2p+1} = 64$ components,
while on the right-hand side there are two scalars, one unit
1-form (with six independent degrees of freedom), plus a generic
2- and 3-form, which also adds up to a total of 64 independent
components. Geometrically, the metric constructed from $\Phi$ and
the 2-form ${\sf B}$ describe the embedding $Spin(7) \times
Spin(7) \subset Spin(7,7)$ while $\varphi$ parameterizes the
conformal factor in $Spin(7,7) \times {\mathbb{R}}^*$. The
parameters $s$ and $c$ can be understood as the sine and cosine of
the angle $\theta$ between the two unit spinors fixed under the
action of each $G_2 \subset Spin(7)$ in the stabilizer.

The next step is to define an even-form $\Box_{\varrho}\varrho =
e^{\sf B} \w *_\Phi \sigma ( e^{-{\sf B}} \w \varrho ) \in
\Lambda^{\rm even} \cong \Lambda^0 \oplus \Lambda^2
\oplus \Lambda^4  \oplus \Lambda^6$, associated with
$\varrho$. The operator $\sigma$ in this expression is the
involution which maps $\sigma(\omega)=-\omega$ for $p$-forms
$\omega$ with $p=1,2 \,{\rm mod}\,4$ and $\sigma(\omega)=\omega$
otherwise. In terms of (\ref{defrho}), the even-form above is
given by
\be \label{defboxrho} \Box_{\varrho}\varrho \; =\; e^{-\varphi}
e^{\sf B} \wedge \left( c -c\, {*_\Phi \Phi} + s {*_\Phi
(\alpha\wedge *_\Phi \Phi )} - s \alpha\wedge \Phi - s *_\Phi
\alpha \right) \; . \ee
The generalized Hitchin functional is then defined as
\be \label{defhi} \int_M \varrho \wedge {\hat \varrho} \; , \ee
where the even-form
\be \hat{\varrho} \; =\;  \sigma(\Box_{\varrho}\varrho ) \; = \;
e^{-\varphi} e^{-{\sf B}} \w *_\Phi \varrho_0 \; , \ee
and we have defined $\varrho_0 = e^\varphi e^{-{\sf B}} \w
\varrho$ in the second equality.

It is easy to verify that
\be \int_M \varrho \wedge {\hat \varrho}  \; =\; \int_M
e^{-2\varphi} \varrho_0 \w *_\Phi \varrho_0 \; =\; \frac{8}{7}
\int_M e^{-2\varphi} \Phi \wedge *_\Phi \Phi \; . \ee
Thus the generalized Hitchin functional looks very similar to the
ordinary one
\footnote{Indeed it can be recast as the ordinary Hitchin functional ${\tilde \Phi} \wedge *_{\tilde \Phi} {\tilde \Phi} = e^{-2\varphi} \Phi \wedge *_\Phi \Phi$ in terms of the rescaled 3-form ${\tilde \Phi} = e^{-6 \varphi /7} \Phi$.}
. In particular, notice that the generalized
functional does not depend on either $\alpha$ or ${\sf B}$ in the non-linear parameterization ({\ref{defrho}}). This
invariance is similar to that found in the case of generalized
Calabi-Yau manifolds {\cite{NH3}}, {\cite{GLW}}, about which more
will be said in the next section.

The first variation of (\ref{defhi}) can be written
\be \label{blu1} 2 \int_M \delta \varrho \wedge \hat{\varrho} \;
=\; 2 \int_M \varrho \wedge \delta \hat{\varrho} \; . \ee
This can be obtained by direct calculation but also follows from
$\int_M \varrho \wedge {\hat \varrho}$ being \lq degree two' in
$\varrho$. Recall that a similar property followed for the
ordinary $G_2$ Hitchin functional from it being a homogeneous
polynomial in $\Phi$ of degree $7/3$
\footnote{Of course things are a bit more subtle in the
generalized case with regard to homogeneity. One finds a
well-defined notion of graded homogeneity exists for terms in
$\varrho$ and ${\hat \varrho}$ if rescaling $\Phi$ with weight 1
is accompanied by rescaling $\alpha$ with weight $1/3$ (all other
fields have weight zero). This structure actually follows from
scaling the constraint $\int_M \alpha\wedge *_\Phi \alpha= \int_M
\frac{1}{7} \Phi \wedge *_\Phi \Phi$.}
. Thus, for variations $\delta \varrho = d \omega$ (for any
$\omega \in \Lambda^{\rm even}$) within a fixed cohomology
class $[ \varrho ] \in H^{\rm odd}(M,{\mathbb{R}})$, the critical points of
the generalized Hitchin functional correspond to generalized $G_2$
manifolds defined by $d \varrho = 0$, $d {\hat \varrho} =0$.

As a quick consistency check of the variation above, notice that
the ${\sf B}$-field variation in $\delta \varrho$ appears in the
form
$$ 2\int_M \delta {\sf B} \wedge \varrho \wedge \hat{\varrho} \; , $$
and one easily verifies that $\varrho\wedge\hat{\varrho}$ has no
5-form component, so that the ${\sf B}$-variation does not
contribute. This agrees with the fact that the generalized Hitchin
functional is independent of ${\sf B}$.


\subsection{Perturbation}

Having found the first variation of the generalized Hitchin
functional (\ref{defhi}), let us now proceed as in the previous
sections and expand $\int_M \varrho \wedge {\hat \varrho}$ to
quadratic order in linear fluctuations of $\varrho$ around a
generalized $G_2$ manifold $M_0$, defined by a fixed background
odd-form ${\bar \varrho}$ that obeys $d {\bar \varrho} =0$ and $d
{\hat {\bar \varrho}} =0$ (where ${\hat{\bar \varrho}} = \sigma (
\Box_{\bar \varrho} {\bar \varrho} )$). To simplify matters, we
will choose this background to be an ordinary $G_2$ holonomy
manifold, i.e. for which $\varphi = {\sf B} = s = 0$ and $\Phi =
\phi$ is the associative 3-form.
%
%

We first expand $\varrho = {\bar \varrho} + \delta \varrho$, and likewise for ${\hat \varrho}$,
to linear order in variations of the parameters $\varphi$, ${\sf
B}$, $\alpha$, $s$ and $\Phi$. One finds the components
\bea
\delta \varrho^1 & = & \delta(s\alpha) \nonumber \\
\delta \varrho^3 & = & \delta\varphi \, \phi - \delta \Phi
-{*(\delta (s\alpha)
\wedge \phi)} \nonumber \\
\delta \varrho^5 & = & - \delta {\sf B} \wedge \phi - \delta
(s\alpha)
\wedge {*\phi} \nonumber \\
\delta \varrho^7 & = & -\frac{1}{7} \delta\varphi \,
\phi\wedge\ast\phi + \frac{1}{3} {*\phi} \w \delta \Phi \; ,
\label{varlist} \eea
for the first order variation of $\varrho$ in (\ref{defrho}), and
\bea
\delta {\hat \varrho}^0 & = & - \delta \varphi  \nonumber \\
\delta {\hat \varrho}^2 & = & -\delta {\sf B} - {*(\delta(s\alpha) \w {*\phi})} \nonumber \\
\delta {\hat \varrho}^4 & = & \delta\varphi \, {*\phi} - \delta( *_\Phi \Phi ) - \delta(s\alpha) \w \phi \nonumber \\
\delta {\hat \varrho}^6 & = & \delta {\sf B} \w {*\phi} + *\delta
(s\alpha )\; , \label{varlist2} \eea
for ${\hat \varrho}$.

In terms of these variations of $\varrho$ and ${\hat \varrho}$,
the generalized Hitchin functional can be expanded as
\be \label{blu2} \int_M \varrho \w {\hat \varrho} \; =\;
\int_{M_0} {\bar \varrho} \w {\hat {\bar \varrho}} + 2 \int_{M_0}
\delta \varrho \wedge {\hat {\bar \varrho}} + \int_{M_0} \delta \varrho \w \delta {\hat
\varrho} \; . \ee
Since we will be interested in only linear variations $\delta \varrho = d\omega$ within a fixed cohomology class, we have not included the additional quadratic term ${1 \over 2} \int_{M_0} \delta^2 \varrho \w {\hat {\bar \varrho}} = {1 \over 2} \int_{M_0} {\bar \varrho}
\w \delta^2 {\hat \varrho}$ which occurs when expanding $\varrho$ and ${\hat \varrho}$ as polynomials in linear variations of the parameters $\varphi$, ${\sf B}$, $\alpha$, $s$ and $\Phi$. It must be stressed that we are assuming it is the linear variations $\delta \varrho$ that describe the degrees of freedom of the quadratic Hitchin action here rather than those of the parameters $\varphi$, ${\sf B}$, $\alpha$, $s$ and $\Phi$. It would be interesting to check this assumption by comparison with the degrees of freedom describing moduli in generalized $G_2$ compactifications of physical string and M-theory.

Plugging the linear variations of $\varrho$ and ${\hat \varrho}$ into the quadratic
term $S_0 = \int_{M_0} \delta \varrho \w \delta {\hat \varrho}$ gives
\bea S_0 &=& \int_{M_0} \delta \Phi \w \delta (
*_\Phi \Phi ) - {8 \over 3} \delta\varphi \, \delta \Phi \w
{*\phi} + {8 \over 7} (\delta \varphi )^2 \phi \w {*\phi} + 8\, \delta (s\alpha) \w {*\delta (s\alpha )} \nonumber \\ [.1in]
&&{\hspace*{.3in}} - 2\, \delta (s\alpha )\w \delta \Phi \w \phi + 4\, \delta (s\alpha) \w \delta {\sf B} \w {*\phi} + \delta {\sf B} \w \delta {\sf B} \w \phi \; .
\label{blu3}
\eea
Notice that demanding linear variations of $\varrho$ has inevitably led to a dependence on the parameters $\alpha$ and ${\sf B}$ in the quadratic part of the generalized Hitchin functional above. This dependence would only be removed by including the term $\int_{M_0} \delta^2 \varrho \w {\hat {\bar \varrho}}$ involving non-linear variations of $\varrho$.

To express $S_0$ in terms of $\delta \varrho$ components, one must invert (\ref{varlist}) to write variations of the parameters in terms of $\delta \varrho$. This gives
\bea \delta \varphi \; \phi & = & {7 \over 4} {\sf P}^3_{\bf 1} (
\delta \varrho^3 ) + {3 \over 4} ( * \delta \varrho^7 ) \, \phi
\nonumber \\ [.1in]
\delta \Phi & = & \frac{3}{4} {\sf
P}^3_{\bf 1}(\delta \varrho^3) -{\sf P}^3_{\bf 7}(\delta
\varrho^3) - {\sf P}^3_{\bf 27}(\delta \varrho^3) + {3 \over 4} (
* \delta \varrho^7 ) \, \phi - {*(\delta \varrho^1 \w \phi )}
\nonumber \\ [.1in]
\delta {\sf B} &=& \left[ 1 - {3 \over 2} {\sf P}^2_{\bf 7} \right] {*\delta \varrho^5} - {1 \over 2} {*(\delta \varrho^1 \w {*\phi})} \; . \label{defvars} \eea
The identities $* {\sf P}^2_{\bf 7} = {1 \over 2} \phi \w {\sf P}^2_{\bf 7}$ and $* {\sf P}^2_{\bf 14} = - \phi \w {\sf P}^2_{\bf 14}$ have been used in deriving the last expression above.

Substituting (\ref{defvars}) into (\ref{blu3}), and using
(\ref{delphi}), implies the quadratic part of the generalized
Hitchin action can be written as
\bea S_0 &=& \int_{M_0} \delta \varrho^3 \w *\left[ \frac{3}{4} {\sf
P}^3_{\bf 1} + {\sf P}^3_{\bf 7}  - {\sf P}^3_{\bf 27} \right]
\delta\varrho^3 -\frac{3}{4} \, \delta \varrho^7 \wedge {*\delta
\varrho^7} -\frac{1}{2} \, {(*\delta\varrho^7)}
\delta \varrho^3 \wedge \ast \phi \nonumber
\\ [.1in] & & + \delta \varrho^5 \w *\left[ \frac{1}{2} {\sf
P}^5_{\bf 7} - {\sf P}^5_{\bf 14} \right]
\delta\varrho^5 - \delta \varrho^1 \w {*\delta \varrho^5} \w {*\phi} -  {1 \over 2}\, \delta \varrho^1 \w {*\delta
\varrho^1}  \nonumber \\ [.1in]
&=& \int_{M_0} \delta \varrho^3 \w *\left[ \frac{4}{3} {\sf
P}^3_{\bf 1} + {\sf P}^3_{\bf 7}  - {\sf P}^3_{\bf 27} \right]
\delta\varrho^3 \nonumber \\
&&-\frac{3}{4} \, \left( \delta \varrho^7  + {1\over 3} \delta \varrho^3 \w {*\phi} \right) \wedge {*\left( \delta \varrho^7  + {1\over 3} \delta \varrho^3 \w {*\phi} \right)} \nonumber
\\ [.1in] & & + \delta \varrho^5 \w *\left[ \frac{1}{2} {\sf
P}^5_{\bf 7} - {\sf P}^5_{\bf 14} \right]
\delta\varrho^5 - \delta \varrho^1 \w {*\delta \varrho^5} \w {*\phi} -  {1 \over 2}\, \delta \varrho^1 \w {*\delta
\varrho^1}   \; . \label{finalvar} \eea
%


\subsection{Quantization}

Let us now take first order variations $\delta \varrho =  d
\omega$ (for any $\omega \in \Lambda^{\rm even}$) within a
fixed cohomology class $[ {\bar \varrho} ] \in H^{\rm odd} (M,{\mathbb{R}})$ of
the background. The quadratic part $S_0$ of the generalized
Hitchin action we have just obtained corresponds to the classical
action to be quantized. If $\omega$ is globally well-defined then
the linear term $\int_{M_0} \delta \varrho \wedge {\hat {\bar
\varrho}}$ in the expansion vanishes since the integrand is a
total derivative. The zeroth order term $\int_{M_0} {\bar \varrho}
\w {\hat {\bar \varrho}} = {8 \over 7} \int_{M_0} \phi \w {*\phi}$
is just proportional to the volume of the background $G_2$
manifold.

Before embarking on this, we must take care that all the
symmetries of the generalized Hitchin action are being accounted
for. Recall that for the ordinary quadratic $G_2$ Hitchin action,
background-preserving diffeomorphisms on $\Phi$ gave rise to the
symmetry under $B \rightarrow B + \iota_v \phi$, for any vector
field $v$ on $M$, in addition to the obvious gauge symmetry under
$B \rightarrow B + d \lambda$. However, it turned out that the
part of $B$ in $\Lambda^2_{\bf 7}$, only contributed a total
derivative to the action and was ignored. Thus, since the
diffeomorphism symmetry only acts on this component of $B$, it was
irrelevant in the quantization of the diffeomorphism-invariant
${\tilde B} \in \Lambda^2_{\bf 14}$ part.

A similar story applies to the generalized $G_2$ Hitchin
functional, but is somewhat more complicated. In this case one has
background-preserving diffeomorphisms plus shifts by exact ${\sf
B}$-fields on $TM \oplus T^* M$ for $\varrho$ which give rise to
the symmetry under $\omega \rightarrow \omega + \iota_v {\bar
\varrho} + \xi \w {\bar \varrho}$, for any vector field $v$ and
1-form $\xi$ on $M$, in addition to the gauge symmetry $\omega
\rightarrow \omega + d \lambda$, for any $\lambda \in \Lambda^{\rm
odd}$. The symmetry $\omega \rightarrow \omega + \iota_v {\bar
\varrho}$ again corresponds to diffeomorphisms of $M$ while
$\omega \rightarrow \omega + \xi \w {\bar \varrho}$ can be understood
as shifting ${\sf B} \rightarrow {\sf B} + d \xi$ by an exact
2-form. These two symmetries also correspond to the subset of
$Spin(7,7)$ transformations that are automorphisms of the Courant
bracket.

Let us first address the latter symmetry. Since ${\bar \varrho} =
-\phi + {1 \over 7} \phi \w {*\phi}$ for the background we have
chosen, the only non-vanishing contribution to $\xi \w {\bar
\varrho}$ is from the 4-form part $- \xi \w \phi$. This
corresponds to a transformation of the component $\omega^4$, where
$\delta \varrho^5 = d \omega^4$. In particular it acts only on the component of $\omega^4$ in the irreducible subspace $\Lambda^4_{\bf 7}$. However, one can check that the two terms involving $\delta \varrho^5 = d\omega^4$ in $S_0$ only contain the component $\omega^4 \in \Lambda^4_{{\bf 27} \oplus {\bf 1}}$, with the ${\bf 7}$ part dropping out as a total derivative, and so this symmetry is redundant.

The only contributions to $\iota_v {\bar \varrho}$ under
diffeomorphisms come from its 2-form and 6-form parts. These
correspond to the transformations $\omega^2 \rightarrow \omega^2 -
\iota_v \phi$ and $\omega^6 \rightarrow \omega^6 + *v^\flat$ of the
components of $\omega$, which certainly do appear in the action
$S_0$ ($v^\flat$ denotes the 1-form dual to vector $v$ in the $\omega^6$
transformation). Notice again that only the part of $\omega^2$ in
$\Lambda^2_{\bf 7}$ transforms under diffeomorphisms. However,
this part of $\omega^2$ does not just give a total derivative
contribution to $S_0$. (Note the factor of $3/4$ in the projector
in square brackets in the first equality in (\ref{finalvar}), relative to the factor
$4/3$ in (\ref{delphi}) that led to a total derivative
contribution for the $\Lambda^2_{\bf 7}$ part.) The trick here, highlighted by the second equality in ({\ref{finalvar}}), is
to observe that although $\omega^6$ and the ${\bf 7}$ part of
$\omega^2$ individually transform under the diffeomorphism
generated by $v$, one can find a particular linear combination of
them
$$C \; :=\; \omega^6 + {1 \over 3} \, \omega^2 \w {*\phi} \; ,$$
that is diffeomorphism-invariant. Therefore, up to total derivatives that we ignore, the action (\ref{finalvar}) can be written purely in terms of the diffeomorphism-invariant fields ${\tilde B} := {\sf P}^2_{\bf 14} \omega^2$, $C$, $D := \omega^0$ and ${\tilde E} := {\sf P}^4_{{\bf 27} \oplus {\bf 1}} \omega^4$, and so we find this symmetry can also be ignored in our quantization.

Thus we are left with only the gauge symmetry under $\delta {\tilde B} = {\sf P}^2_{\bf 14} d \lambda$, $\delta C = d \mu$ and $\delta {\tilde E} = {\sf P}^4_{{\bf 27} \oplus {\bf 1}} d \nu$ in $S_0$, for any $\lambda \in \Lambda^1$, $\mu \in
\Lambda^5$ and $\nu \in \Lambda^3_{{\bf 27}\oplus {\bf 7}}$. The required prefactor ${\sf P}^4_{{\bf 27} \oplus {\bf 1}}$ in the gauge transformation for ${\tilde E}$ projects out any singlet component of $\nu$ identically.

Before going on to consider the partition function for $S_0$, it will be convenient to illustrate how a field redefinition involving a shift by $D$ of the singlet part of ${\tilde E}$ can be used to remove the term $\int_{M_0} \delta \varrho^1 \w {*\delta \varrho^5} \w {*\phi}$ from the action. This works by first noting the identity
\be
{1 \over 2} | {\sf P}^5_{\bf 7} \delta \varrho^5 - \delta \varrho^1 \w {*\phi} |^2 -2\, | \delta \varrho^1 |^2 \; =\; {1 \over 2} | {\sf P}^5_{\bf 7} \delta \varrho^5 |^2 - \delta \varrho^1 \w {*\delta \varrho^5} \w {*\phi} - {1 \over 2} | \delta \varrho^1 |^2 \; ,
\ee
which follows using $| \xi \w {*\phi} |^2 = 3\, | \xi |^2$ for any 1-form $\xi$. In addition, one can check that ${\sf P}^5_{\bf 14} \delta \varrho^5 = {\sf P}^5_{\bf 14} d {\tilde E}$ projects out the singlet part of ${\tilde E}$. Thus by redefining the singlet part ${\sf P}^4_{\bf 1} {\tilde E} \rightarrow {\sf P}^4_{\bf 1} {\tilde E} - D \, {*\phi}$ one can rewrite the action (\ref{finalvar}) more conveniently as
\bea \label{finalvar2} S_0  &=& \int_{M_0} d{\tilde B} \w
* \left( 2\, {\sf P}^3_{\bf 7} - 1 \right) d {\tilde B} -\frac{3}{4}
\, dC \wedge {*dC}  \nonumber \\ [.1in] &&{\hspace*{.3in}} -2 \,
dD \w {*dD} + d{\tilde E} \w * \left( {3 \over 2}\, {\sf P}^5_{\bf
7} - 1 \right) d {\tilde E} \; ,
\eea
where ${\tilde E}$ is now the redefined field. This redefinition has therefore diagonalized the classical
action. Notice that the quadratic generalized Hitchin action
(\ref{finalvar2}) contains the ordinary quadratic Hitchin action we quantized in sections 3 and 4. In addition there are the decoupled actions for a free 6-form $C$ and scalar $D$, plus the somewhat more complicated action for the 4-form ${\tilde E} \in \Lambda^4_{{\bf 27} \oplus {\bf 1}}$. The BV quantizations of $C$ and ${\tilde E}$ are detailed in appendices D and E respectively.


\subsection{Partition function}

To obtain the 1-loop partition function for the generalized
Hitchin action, we can just multiply the 1-loop partition function
found previously for the ordinary $G_2$ Hitchin action with those
for the decoupled fields $C$, $D$ and ${\tilde E}$.

The action for the scalar $D$ is non-degenerate and so its
partition function is simply $Z_0 = ({\mbox{det}}\, \Delta^0_{\bf
1} )^{-1/2}$. The partition function for the 6-form $C$ was
calculated in appendix D and found to equal the reciprocal of the
Ray-Singer torsion of the background $G_2$ manifold, $Z_6 =
I_{RS}^{-1}$. The calculation in appendix E also yielded $Z_4^{{\bf 27} \oplus {\bf 1}} = I_{RS}^{-1}$ for ${\tilde E}$. Using the expression $Z = Z_2^{\bf 14}$ in (\ref{partf2}) for the
partition function of the ordinary quadratic Hitchin action, the
1-loop partition function $Z_{gen}$ for the generalized Hitchin
action can be written
\bea Z_{gen} &=& Z_0 Z_6 Z_4^{{\bf 27} \oplus {\bf 1}} Z_2^{\bf 14} \nonumber \\
&=& \left[ ({\mbox{det}}\, \Delta^0_{\bf 1} )^{-1/2} \right]
\times \left[ ({\mbox{det}}\, \Delta^3 )^{1/2}
( {\mbox{det}}\, \Delta^2 )^{-3/2} ({\mbox{det}}\, \Delta^1 )^{5/2} ({\mbox{det}}\, \Delta^0 )^{-7/2} \right]  \nonumber \\
[.1in] && \times \left[ ({\mbox{det}}\, \Delta^4_{{\bf 27} \oplus {\bf 1}} )^{-1/2} ({\mbox{det}}\, \Delta^3_{{\bf 27} \oplus {\bf 7}} ) ({\mbox{det}}\, \Delta^2 )^{-3/2} ({\mbox{det}}\, \Delta^1  )^{2} ({\mbox{det}}\, \Delta^0 )^{-5/2} \right] \nonumber \\ [.1in]
&& \times \left[ ({\mbox{det}}\, \Delta^2_{\bf 14} )^{-1/2}
( {\mbox{det}}\, \Delta^1_{\bf 7} ) ({\mbox{det}}\, \Delta^0_{\bf
1} )^{-3/2} \right] \nonumber \\ [.1in] &=& ({\mbox{det}}\,
\Delta_{\bf 1} )^{-8} ({\mbox{det}}\, \Delta_{\bf 7} )^{4} (
{\mbox{det}}\, \Delta_{\bf 14} )^{-7/2} ({\mbox{det}}\, \Delta_{\bf
27} ) \; . \label{genpart} \eea
The second equality follows using Hodge duality ${\mbox{det}}\,
\Delta^p = {\mbox{det}}\, \Delta^{7-p}$ to simplify $Z_6$. The
final equality uses orthogonality of Laplacians acting on $G_2$
irreps to write ${\mbox{det}}\, \Delta^{3} = ( {\mbox{det}}\,
\Delta^{3}_{\bf 1} ) ( {\mbox{det}}\, \Delta^{3}_{\bf 7} ) (
{\mbox{det}}\, \Delta^{3}_{\bf 27} )$, ${\mbox{det}}\, \Delta^{2}
= ( {\mbox{det}}\, \Delta^{2}_{\bf 7} ) ( {\mbox{det}}\,
\Delta^{2}_{\bf 14} )$ and also the various $G_2$ isomorphisms to
relate ${\mbox{det}}\, \Delta_{\bf 1} = {\mbox{det}}\,
\Delta^{0}_{\bf 1} = {\mbox{det}}\, \Delta^{3}_{\bf 1}$,
${\mbox{det}}\, \Delta_{\bf 7} = {\mbox{det}}\, \Delta^{1}_{\bf 7}
= {\mbox{det}}\, \Delta^{2}_{\bf 7} = {\mbox{det}}\,
\Delta^{3}_{\bf 7}$ (${\mbox{det}}\, \Delta_{\bf 14} =
{\mbox{det}}\, \Delta^{2}_{\bf 14}$ and ${\mbox{det}}\,
\Delta_{\bf 27} = {\mbox{det}}\, \Delta^{3}_{\bf 27}$).


\section{Topological $G_2$ string at one loop}
\setcounter{equation}{0}

The genus-one free energy for a closed string theory is given by
\begin{equation}
F_1 \; =\; \int \frac{d\tau d\bar{\tau}}{\tau_2} \, \textrm{Tr}
\left( (-1)^F F_L F_R e^{2\pi \tau i H_L - 2 \pi \bar{\tau} i H_R}
\right) \; .
\end{equation}
If we treat this as an integral over the upper half plane, rather
than the fundamental domain of the torus complex structure
$\tau = \tau_1 + i \tau_2$, then it simplifies to
\bea F_1 &=&  \delta (H_L - H_R) \int \frac{d\tau_2}{\tau_2} \,
\textrm{Tr} \left( (-1)^F F_L F_R e^{2\pi i \tau_2 (H_L + H_R)}
\right) \nonumber \\ [.1in] &=&  \delta(H_L - H_R) \, {\rm{log}}
\left[  \prod_{F_L, F_R} \textrm{det} (2\pi(H_L + H_R))^{(-1)^F
F_L F_R}  \right] \; . \eea
Here $F_L$ and $F_R$ are the right- and left-handed fermion number
operators and $F = F_L + F_R$.  To evaluate this expression we
need to know how the total Hamiltonian $H_L + H_R$ acts on a
general state $A_{M_1... M_i N_1 ... N_j}(X) \psi_L^{M_1}
...\psi_L^{M_i} \psi_R^{N_1} ... \psi_R^{N_j}$ in the Hilbert
space of closed $G_2$ string states.  Recall from
{\cite{g2string}} that the left- and right-moving sectors of the
worldsheet each span a copy of the ${\check D}$ complex in
({\ref{g2dol}}) in the $G_2$ string Hilbert space, with the left-
and right-moving BRST operators $Q_L$ and $Q_R$ identified with
${\check D}$ acting on each of these two copies.

The Hamiltonians in the left- and right-moving sectors are $H_L =
\{Q_L^\dagger, Q_L\}$ and $H_R = \{Q_R^\dagger, Q_R\}$.  The
$G_2$-irreducible $p$-form spaces $\Lambda^p_{\bf n}$ in the same
$G_2$ irrep ${\bf n}$ are isomorphic for different values of $p$,
and the action of the operators $H_L$ and $H_R$ on $\Lambda^p_{\bf
n}$ depends only on the dimension of the $G_2$ irrep. Thus we need
only determine how these operators act on the tensor products of
the {\bf 1} and {\bf 7} (i.e. states with 0 or 1 fermion in the
left and right sectors), since the action of $H_L + H_R$ on the
other states will follow from this. This is done in appendix F
where the Hamiltonian is found to act as the Laplacian operator
$\Delta_{{\bf 7} \otimes {\bf 7}} = \Delta^2_{\bf 14} +
\Delta^2_{\bf 7} + \Delta^3_{\bf 27} + \Delta^3_{\bf 1}$ on states
in ${\bf 7} \otimes {\bf 7}$, $\Delta_{\bf 7} = \Delta^1_{\bf 7}$
on states in ${\bf 7} \otimes {\bf 1} \cong {\bf 1} \otimes {\bf
7}$ and $\Delta_{\bf 1} = \Delta^0_{\bf 1}$ on states in ${\bf 1}
\otimes {\bf 1}$.

Having obtained the action of $H_L + H_R$ on the $G_2$ string
spectrum, we are now prepared to evaluate the one-loop result
\be \log \left[  \prod_{F_L, F_R} \textrm{det} (2\pi(H_L +
H_R))^{(-1)^F F_L F_R} \right] \; . \ee
Recall that the fermion numbers $F_L$ and $F_R$ run from 0 to 3,
each labeling elements of the respective left/right copy of the
${\check D}$ complex $\Lambda^0_{\bf 1} \rightarrow \Lambda^1_{\bf
7} \rightarrow \Lambda^2_{\bf 7} \rightarrow \Lambda^3_{\bf 1}$.
The relevant $G_2$ irreps are thus {\bf 1}, {\bf 7}, {\bf 7}, and
{\bf 1} for 0, 1, 2, and 3 respectively.  To compute all the
contributions in the product above we determine $(H_L +
H_R)^{(-1)^F F_L F_R}$ for all values of $F_L$ and $F_R$ in the
table below

\begin{center}
\begin{tabular}{|c | c | c | c |c |}
\hline
$F_L$  / $F_R$  & 0 &  1 & 2 & 3 \\
\hline
0  & $(\Delta_{\bf 1})^0$ &  $(\Delta_{\bf 7})^0$ & $(\Delta_{\bf 7})^0$ & $(\Delta_{\bf 1})^0$ \\
\hline
1  & $(\Delta_{\bf 7})^0$ &  $(\Delta_{{\bf 7} \otimes {\bf 7}})^1$ & $(\Delta_{{\bf 7} \otimes {\bf 7}})^{-2}$ & $(\Delta_{\bf 7})^3$ \\
\hline
2  & $(\Delta_{\bf 7})^0$ &  $(\Delta_{{\bf 7} \otimes {\bf 7}})^{-2}$ & $(\Delta_{{\bf 7} \otimes {\bf 7}})^{4}$ & $(\Delta_{\bf 7})^{-6}$ \\
\hline
3  & $(\Delta_{\bf 1})^0$ &  $(\Delta_{\bf 7})^3$ & $(\Delta_{\bf 7})^{-6}$ & $(\Delta_{\bf 1})^9$ \\
\hline
 \end{tabular}
\end{center}

Combining all these contributions gives $( \det \, \Delta_{{\bf 7}
\otimes {\bf 7}} ) ( \det \, \Delta_{\bf 7} )^{-6} ( \det \,
\Delta_{\bf 1} )^9$ which can be further simplified by decomposing
the first determinant in terms of Laplacians acting on $G_2$
irreps. That is $\det \, \Delta_{{\bf 7} \otimes {\bf 7}} = (
\det\, \Delta_{\bf 14} ) ( \det\, \Delta_{\bf 7} ) ( \det\,
\Delta_{\bf 27} ) ( \det\, \Delta_{\bf 1} )$ because composition
of any two different irreducible Laplacians in $\Delta^2_{\bf 14}
+ \Delta^2_{\bf 7} + \Delta^3_{\bf 27} + \Delta^3_{\bf 1}$
vanishes as a result of orthogonality of the $G_2$ projectors.
Thus we have
\bea && ( \det \, \Delta_{{\bf 7} \otimes {\bf 7}} ) ( \det \,
\Delta_{\bf 7} )^{-6} ( \det \, \Delta_{\bf 1} )^9 \; =\;\nonumber \\
& & \quad \quad ( \det
\, \Delta_{\bf 1} )^{10} ( \det \, \Delta_{\bf 7} )^{-5} ( \det \,
\Delta_{\bf 14} ) ( \det \, \Delta_{\bf 27} )   \; . \eea
It will be convenient to normalize such that the 1-loop partition
function for the $G_2$ string is $Z_{string} = {\rm{exp}} \left(
-{1 \over 2} F_1 \right)$, so that
\be Z_{string} \; =\; ({\mbox{det}}\, \Delta_{\bf 1} )^{-5}
({\mbox{det}}\, \Delta_{\bf 7} )^{5/2} ( {\mbox{det}}\,
\Delta_{\bf 14} )^{-1/2} ({\mbox{det}}\, \Delta_{\bf 27} )^{-1/2}
\; . \label{stringpart} \ee

Thus we conclude that $Z_{string} \neq Z_{gen}$ but their relation will be explained in more detail in section 8.


\section{Dimensional reduction} \label{dimred}
\setcounter{equation}{0}

Let us now consider a special $G_2$ background of the form $M_0 =
CY_3 \times S^1$ and compactify the classical quadratic Hitchin
functional ({\ref{Scl}}) on the circle. The purpose of this
reduction will be comparison with previous work on quantizing
Hitchin functionals in 6 dimensions.

The background and perturbation reduce to
\bea
\phi &=& k \w dt + {\rho} \nonumber \\
*\phi &=& {\hat \rho} \w dt + {1 \over 2} \, k\w k \nonumber \\
{\tilde B} &=& A \w dt + b \; , \label{bakred} \eea
where $k$ is the K\"{a}hler form, $\rho$ and ${\hat \rho}$ are the
real and imaginary parts of the holomorphic $(3,0)$-form on $CY_3$
and $t$ is the $S^1$ coordinate. Recall that ${\tilde B} \in
\Lambda^2_{\bf 14}$ and the constraint $\phi^{MNP} {\tilde B}_{NP}
=0$ implies the 1-form $A$ and 2-form $b$ in six dimensions are
not linearly independent. In particular, the reduction of this
constraint implies
$$
\rho_{mnp} \, b^{np} + 2\, k_{mn} A^n \; =\; 0 \; , \quad k_{mn}
b^{mn} \; =\; 0 \; .
$$
Since $\Lambda^2 = \Lambda^{20} \oplus \Lambda^{11} \oplus
\Lambda^{02} = \Lambda^2_{\bf 3} \oplus \Lambda^2_{\bf 8} \oplus
\Lambda^2_{\bf 1} \oplus \Lambda^2_{\bar {\bf 3}}$ in terms of
$SU(3)$ representations, the equations above tell us that $b$ has
no ${\bf 1}$ singlet part and its ${\bf 3} \oplus {\bar {\bf 3}} \cong {\bf 6}$
vector part is proportional to $A$. It will prove more convenient
to remove this dependence by describing the reduction in terms of
the redefined field
\be {\tilde b}_{mn} \; =\; b_{mn} + \frac{1}{2} {\hat \rho}_{mnp}
A^p \; , \ee
which obeys $\rho_{mnp} {\tilde b}^{np} =0$, $k_{mn} {\tilde
b}^{mn} =0$ and so ${\tilde b} \in \Lambda^2_{\bf 8}$.

The quadratic terms in the $G_2$ Hitchin action reduce to
\bea | \phi_{MNP} {\tilde H}^{MNP} |^2 &=& 9\, | k_{mn} F^{mn} |^2
+ 6\, k_{ab} F^{ab} \rho_{mnp} h^{mnp} + | \rho_{mnp} h^{mnp} |^2
 \nonumber \\ [.1in]
| {*\phi}_{MNPQ} {\tilde H}^{NPQ} |^2 &=& 18\, | F_{mn} |^2 - 18\, F_{mn} F_{pq} k^{mp} k^{nq} - 18\, \rho^{mab} F_{ab} k^{cd} h_{mcd} \nonumber \\
&&+ | {\hat \rho}_{mnp} h^{mnp} |^2 + 9\, h^m_{\;\;\; ab} h_{mcd}
k^{ab} k^{cd} \nonumber \\ [.1in] | {\tilde H}_{MNP} |^2 &=& 3\, |
F_{mn} |^2 + | h_{mnp} |^2 \; , \eea
where $F=dA$ and $h=db$. Some algebraic identities for products of the background Calabi-Yau data $k$ and $\rho$ have been used, which follow by substituting ({\ref{bakred}}) into the $G_2$ identities in appendix A. The first term vanishes due to
$\phi^{MNP} {\tilde B}_{NP} =0$. The second and third terms can be
expressed in terms of ${\tilde h} = d {\tilde b}$ and $F=dA$ and
simplified.

The final result is that
\bea | {\tilde H}_{MNP} |^2 - \frac{1}{12} | {*\phi}_{MNPQ}
{\tilde H}^{NPQ} |^2 &=& | {\tilde h}_{mnp} |^2 - \frac{3}{4} |
{\tilde h}_{mnp} k^{np} |^2 \nonumber \\ [.1in] && -3 \left(
{\tilde h}_{mnp} {\hat \rho}^{mnq} \, F^p_{\;\;\, q} - \frac{1}{4}
{\tilde h}_{mab} k^{ab} \rho^{mcd} F_{cd} \right) \nonumber \\
[.1in] &&+ \frac{9}{2} \left( | F_{mn} |^2 - \frac{1}{8} |
\rho_{mnp} F^{np} |^2 \right) \; , \label{quadred} \eea
up to total derivatives which we ignore. The integral of the left
hand side being proportional to the quadratic Hitchin action
({\ref{Scl}}) in 7 dimensions.

It is worth noting that this reduced action has quite a subtle
gauge symmetry arising from reduction of the symmetry under
$\delta {\tilde B} = {\sf P}^2_{\bf 14} d \lambda$ in 7
dimensions. It is invariant under the transformations
\be \delta {\tilde b} \; =\; {\sf P}^2_{\bf 8} d \lambda \; ,
\quad \delta A_m \; =\; -\frac{1}{3} {\hat \rho}_{mnp} \partial^n
\lambda^p + \frac{2}{3} \partial_m \alpha \; , \ee
where $\lambda$ and $\alpha$ are a 1-form and a scalar in 6
dimensions. The $\alpha$ transformation leaving $F=dA$ invariant
is the usual gauge symmetry but notice that $A$ also transforms
under the canonical gauge transformation for the 2-form ${\tilde
b}$. It will be convenient to introduce the dual variable $\beta_{mn} = - 3\, \rho_{mnp} A^p \in \Lambda^2_{\bf 6}$ to the gauge field $A_m$, which has the gauge transformation $\delta \beta_{mn} = \rho_{mnp} {\hat \rho}^{pab} \partial_a \lambda_b -2\, \rho_{mnp} \partial^p \alpha$. In terms of this dual field, the Lagrangian ({\ref{quadred}}) becomes
\bea
&& | {\tilde h}_{mnp} |^2 - \frac{3}{4} | {\tilde h}_{mnp} k^{np} |^2  - \frac{3}{2} {\tilde h}_{mnp} k^{np} \partial_q \beta^{mq} + \frac{3}{4} | \partial^n \beta_{mn} |^2 \nonumber \\ [.1in]
&&+ \frac{1}{16} | {\hat \rho}_{mnp} \partial^m \beta^{np} |^2 \; , \label{quadred2} \eea
up to total derivatives
\footnote{The identities $k_m^{\;\;\; p} {\tilde b}_{pn} = k_n^{\;\;\; p} {\tilde b}_{pm}$ and $k_m^{\;\;\; p} \beta_{pn} = - k_n^{\;\;\; p} \beta_{pm}$, which hold for any ${\tilde b} \in \Lambda^2_{\bf 8}$ and $\beta \in \Lambda^2_{\bf 6}$, have been used. In addition, it is helpful in deriving ({\ref{quadred2}}) to use the identity $2\, | F_{mn} |^2 = | \rho_{mnp} F^{np} |^2 + | k^{mn} F_{mn} |^2$ (up to total derivatives) and that $\rho_{mnp} F^{np} = -\frac{2}{3} \partial^n \beta_{mn}$ and $k^{mn} F_{mn} = \frac{1}{6} {\hat \rho}_{mnp} \partial^m \beta^{np}$.}
. Notice that $\partial^n \beta_{mn}$ is invariant under the $\alpha$ part of the gauge transformation while ${\hat \rho}_{mnp} \partial^m \beta^{np}$ is invariant under the full gauge transformation of $\beta$. $\partial^n \beta_{mn}$ transforms non-trivially under the $\lambda$ part of the gauge transformation but the integral of the first line in ({\ref{quadred2}}) is fully gauge-invariant.

We are now prepared to compare these results with the analysis of Pestun and Witten {\cite{PW}}.


\subsection{Comparison with Pestun-Witten}

The field $b$ we obtain from dimensional reduction of the $G_2$
Hitchin functional may look reminiscent of the 2-form appearing in
the quantization of the quadratic Hitchin action for a stable
3-form in 6 dimensions {\cite{PW}}. Indeed one might expect the
results of Pestun and Witten to follow as some kind of consistent
truncation of the reduction of the $G_2$ theory. After all, the
variations $\rho \rightarrow \rho + db\, '$ considered in
{\cite{PW}} form a subset of the ones $\phi \rightarrow \phi + dB$
we have used in 7 dimensions (within which $k$ is invariant).
We will now show that this is indeed the case, though the relationship between the
quadratic actions is not quite so straightforward.

Under variations $\rho \rightarrow \rho + db\, '$, the quadratic
part of the Hitchin functional $\int \rho \w {\hat \rho}$ for a
stable 3-form in 6 dimensions is proportional to
\be \label{6quad} \int d^6\, x \; \left[ | h'_{mnp} |^2 -
\frac{3}{2} | h'_{mnp} k^{np} |^2 - \frac{1}{12} | h'_{mnp} \,
\rho^{mnp} |^2 - \frac{1}{12} | h'_{mnp} \, {\hat \rho}^{mnp} |^2
\right] \; , \ee
where $h' = d b\, '$. This is just a rewriting of the classical
action in equation (2.11) of {\cite{PW}} in real coordinates, and
equals $6 \int h' \w J h'$, where
$$
J_{mnp}^{\;\;\;\;\;\;\;\, qrs} \; =\; -\frac{1}{6}
\epsilon_{mnp}^{\;\;\;\;\;\;\;\, qrs} + \frac{3}{2} k_{[mn}
k_{p]}^{\;\; [q} k^{rs]} - \frac{1}{12} {\hat \rho}_{mnp}
\rho^{qrs} + \frac{1}{12} \rho_{mnp} {\hat \rho}^{qrs} \; ,
$$
defines the action of the complex structure of the background on
3-forms (and indeed obeys $J^2=-1$ and $J \Omega = i \Omega$,
$\Omega = \rho + i {\hat \rho}$).

Let us now decompose $b_{mn}' = {\hat b}_{mn} + \frac{1}{4}
\rho_{mnp} a^p$, where $a_m = \rho_{mnp} b\,'^{\, np}$ and ${\hat
b} \in \Lambda^2_{\bf 8} \oplus \Lambda^2_{\bf 1}$ ($a$ and ${\hat
b}$ correspond to $b_{20}+b_{02}$ and $b_{11}$ in {\cite{PW}} in
complex coordinates). Just as in equation (2.11) of {\cite{PW}},
one finds that all terms involving $a$ drop out of ({\ref{6quad}})
as total derivatives, making background-preserving diffeomorphisms
a redundant symmetry of this action. The resulting Lagrangian is
proportional to
\be \label{6quad2} | {\hat h}_{mnp} |^2 - \frac{3}{2} | {\hat
h}_{mnp} k^{np} |^2 \; , \ee
where ${\hat h} = d{\hat b}$, and its integral is invariant under
the gauge transformation $\delta {\hat b} = {\sf P}^2_{{\bf 8}
\oplus {\bf 1}} d \lambda$, for any 1-form $\lambda$. (This fact
is more obvious in complex coordinates where the Lagrangian above
is $\partial b_{11} \w {\bar \partial} b_{11}$ and $\delta b_{11}
=
\partial \lambda_{01} + {\bar
\partial} \lambda_{10}$.)

We can further decompose the 2-form ${\hat b}_{mn} = {\tilde
b}_{mn} + \frac{1}{6} k_{mn} \varphi$ into irreducible
representations of $SU(3)$, where $\varphi = k^{mn} {\hat b}_{mn}$
is its singlet part and ${\tilde b} \in \Lambda^2_{\bf 8}$ is its
primitive component in the adjoint of $SU(3)$, that we would like to relate to the 2-form
gauge field appearing in ({\ref{quadred}}). Under this
decomposition, ({\ref{6quad2}}) reduces to
\bea | {\hat h}_{mnp} |^2 - \frac{3}{2} | {\hat h}_{mnp} k^{np}
|^2 &=& | {\tilde h}_{mnp} |^2 - \frac{3}{2} | {\tilde h}_{mnp}
k^{np} |^2 \nonumber \\ [.1in] && - {\tilde h}_{mnp} k^{np}
\partial^m \varphi - \frac{1}{3} | \partial_m \varphi |^2 \; , \label{6quad3} \eea
and is invariant under the gauge transformations $\delta {\tilde
b} = {\sf P}^2_{\bf 8} d \lambda$, $\delta \varphi = 2 \, k^{mn}
\partial_m \lambda_n$.

Notice that naively setting $\varphi =0$ would not identify this
action with the first line of ({\ref{quadred}}). This could have
been anticipated though since neither $\varphi =0$ nor $F_{mn} =0$ are
gauge-invariant equations. A better strategy is to integrate out
$\varphi$. The gauge-invariant equation of motion for $\varphi$ is
\be \frac{2}{3} \, \square \varphi \; =\; - \partial^m ( {\tilde
h}_{mnp} k^{np} ) \; . \ee
This implies the equation
\be \label{const} \frac{2}{3} \, \partial_m \varphi \; =\;
- {\tilde h}_{mnp} k^{np} + \partial^n \beta_{mn} \; , \ee
where the coexact term involves some locally-defined 2-form $\beta$. The equation above is only gauge-invariant provided $\delta \beta_{mn} = \rho_{mnp} {\hat \rho}^{pab} \partial_a \lambda_b + \rho_{mnp} \partial^p \gamma + {\hat \rho}_{mnp} \partial^p \varepsilon$, for any scalars $\gamma$ and $\varepsilon$. Thus we can identify $\beta \in \Lambda^2_{\bf 6}$ in the coexact term above with the dual variable to $A$ introduced in the previous subsection (provided we set $\gamma = -2\alpha$ and $\varepsilon =0$).

Substituting the equation above into ({\ref{6quad3}}) implies the Pestun-Witten
Lagrangian becomes
\be | {\tilde h}_{mnp} |^2 - \frac{3}{4} | {\tilde h}_{mnp} k^{np}
|^2 - \frac{3}{2} {\tilde h}_{mnp} k^{np} \partial_q \beta^{mq} + \frac{3}{4} | \partial^n \beta_{mn} |^2 \; , \ee
which agrees with the first line of ({\ref{quadred2}}). The absence of the second line of ({\ref{quadred2}}) in the Lagrangian above is due to the extra scalar gauge symmetry under $\delta \beta_{mn} = {\hat \rho}_{mnp} \partial^p \varepsilon$ in the Pestun-Witten theory, which does not arise from reduction of the $G_2$ theory in seven dimensions. However, the second line of ({\ref{quadred2}}) has a nice interpretation from gauge-fixing the extra $\varepsilon$ symmetry in the Lagrangian above. That is, under $\delta \beta_{mn} = {\hat \rho}_{mnp} \partial^p \varepsilon$, the dual 1-form gauge field $k_{mn} A^n$ has the canonical gauge transformation $\partial_m \varepsilon$. Thus the Lorentz gauge-fixing term for this symmetry is proportional to $| \partial^m ( k_{mn} A^n )|^2$ which is exactly the square of $k^{mn} F_{mn} = {1 \over 6} {\hat \rho}_{mnp} \partial^m \beta^{np}$ appearing in the second line of ({\ref{quadred2}}).

Thus we have found agreement between the local degrees of freedom arising from the reduction of the $G_2$ theory and the Pestun-Witten theory describing variations of a stable 3-form in six dimensions. This may seem somewhat surprising since we were allowing variations of both $k$ and $\rho$ in the reduced theory. Indeed the premise of topological M-theory {\cite{M}} is that classically the $G_2$ Hitchin functional should encapsulate {\emph{both}} K\"{a}hler and complex structure deformations of the A- and B-models in six dimensions. Thus, in addition to the Pestun-Witten theory, we might have expected the quadratic action for a stable 2-form, that is related to the quantum foam description of the A-model {\cite{qf1,qf2}}, from the reduction. However, such a quadratic action would be proportional to $\int k\w F\w F$ and so the Lagrangian corresponds to a locally-defined total derivative. For general Calabi-Yau backgrounds this term corresponds to the non-trivial integral second Chern class of the $U(1)$ gauge bundle with curvature $F$. However, in the topologically trivial case we have considered, such terms have been dropped. It would be interesting to understand the global topological structure of the reduced theory in more detail, but this would require a more refined analysis than we are attempting here.


\subsection{Dimensional reduction of generalized $G_2$ theory}

Having related the dimensional reduction of the quadratic $G_2$
Hitchin functional to the corresponding quantity in six dimensions
calculated in {\cite{PW}}, we will now examine the reduction of
the generalized $G_2$ theory and its relation to the
extended Hitchin functional used in {\cite{PW}}. To do this it
will be helpful to begin with a brief review of generalized
Calabi-Yau manifolds (see {\cite{NH3}}, {\cite{GLW}} for more
details).


\subsubsection{Generalized Calabi-Yau manifolds}

The critical points of the generalized Hitchin functional in six
dimensions correspond to six-manifolds $N$ with
{\emph{generalized $SU(3)$ structure}}, so called
{\emph{generalized Calabi-Yau}} manifolds {\cite{NH3}}. The
structure group $Spin(6,6)$ of $TN \oplus T^* N$ here is reduced
to an $SU(3)\times SU(3)$ subgroup in the following way. Under the
action of the conformal structure group $Spin(6,6) \times {\mathbb
R}^*$ in six dimensions, the stabilizer of a generic form of
either odd or even degree is $SU(3,3)$. When acting on
complex-valued odd/even-forms, the conformal structure group is
complexified to $Spin(12,{\mathbb{C}}) \times {\mathbb{C}}$ and
its orbits correspond to the subspaces $\Lambda^{30} \oplus
\Lambda^{21} \oplus \Lambda^{10} \oplus \Lambda^{32} \subset
\Lambda^{\rm odd} \otimes {\mathbb{C}}$ and ${\mbox{exp}} (
\Lambda^0 \otimes {\mathbb{C}} \oplus \Lambda^2 \otimes
{\mathbb{C}} ) \subset \Lambda^{\rm even} \otimes {\mathbb{C}}$ on
$N$, each of which are fixed by an $SU(3,3)$ subgroup. Both these
$SU(3,3)$-invariant orbits are 32-dimensional and, as vector
spaces, are isomorphic to the real form subspaces $\Lambda^{\rm
odd/even} \subset \Lambda^{\rm odd/even} \otimes {\mathbb{C}}$.
Given two generic stable forms of odd and even degrees, a
different $SU(3,3)$ stabilizes each of them and it is only a
common $SU(3)\times SU(3)$ subgroup that can fix them both
simultaneously. An odd- and even-form which are simultaneously
stabilized by $SU(3) \times SU(3)$ in this way are said to be
{\emph{compatible}}. The existence of a stable odd- and even-form
which are compatible defines a generalized Calabi-Yau structure
\footnote{This is similar to the situation for ordinary Calabi-Yau
structures in six dimensions, where the stable 2-form $k$ (fixed
by $Sp(3,{\mathbb{R}}) \subset GL(6,{\mathbb{R}})$) and 3-form
$\rho$ (fixed by $SL(3,{\mathbb{C}}) \subset GL(6,{\mathbb{R}})$)
can only be simultaneously fixed by a common $SU(3)$ subgroup.
These two forms are compatible if $k\w \rho =0$.}
. As noted in equation (2.102) of {\cite{GLW}}, any two
stable forms $\chi_- \in \Lambda^{\rm odd}$ and $\chi_+ \in
\Lambda^{\rm even}$ are guaranteed to be compatible provided they
solve
\bea \langle (v+\xi )\cdot \chi_- , \chi_+ \rangle &=&  \iota_v
\chi_-^1 \w \chi_+^6 - ( \iota_v \chi_-^3 + \xi\w \chi_-^1 ) \w
\chi_+^4 \nonumber \\ [.1in] &&+ ( \iota_v \chi_-^5 + \xi\w
\chi_-^3 ) \w \chi_+^2 - \xi\w \chi_-^5 \w \chi_+^0 \nonumber \\
[.1in] &=& 0 \; , \label{gencom} \eea
for any vector $v$ and 1-form $\xi$ on $N$. Since $v+\xi$ transforms as a vector under $Spin(6,6)$, this condition is clearly necessary due to the absence any singlets in the vector decomposition under $SU(3)\times SU(3) \subset Spin(6,6)$. The
operator $(v+\xi )\cdot = \iota_v + \xi \wedge$ gives the action
of the Clifford algebra on odd/even-forms (understood as
Majorana-Weyl spinors of $Spin(6,6)$). The bilinear map $\Lambda^{\rm odd/even} \times \Lambda^{\rm odd/even}
\rightarrow \Lambda^6$
$$\langle \omega^{\rm odd}, \chi^{\rm odd} \rangle = - \omega^1 \w \chi^5 +
\omega^3 \w \chi^3 - \omega^5 \w \chi^1$$
$$\langle \omega^{\rm
even}, \chi^{\rm even} \rangle = \omega^0 \w \chi^6 - \omega^2 \w
\chi^4 + \omega^4 \w \chi^2 - \omega^6 \w \chi^0 \; ,$$
called the {\emph{Mukai pairing}}, represents the inner product of the isomorphic $Spin(6,6)$ chiral spinors.

A special case where the generalized Calabi-Yau structure reduces
to an ordinary one is when the stable odd-form has no 1-form and
5-form components but the even-form is generic. The odd-form is
then a 3-form $\rho$, stabilized by $SL(3,{\mathbb{C}})$. If we
call the complex 2-form ${\sf b}+ik$ in the even-form orbit then
the 12 generalized compatibility equations reduce to $\rho \w k =
\rho \w {\sf b} =0$ and solutions define an ordinary $SU(3)$
structure (corresponding to the common subgroup of odd/even-form
stabilizers $SL(3,{\mathbb{C}})$ and $SU(3,3)$).


\subsubsection{Reduction of generalized $G_2$ theory}

The parameterization given in \cite{wi1,wi2,wi3} for the stable
odd-form $\varrho$ ({\ref{defrho}}) we used in seven dimensions is
convenient for the reduction since, in an orthonormal frame, the
1-form $\alpha$ defines the direction orthogonal to which the
generalized Calabi-Yau structure is contained. Thus we will
decompose the data with respect to the direction defined by
$\alpha$ as
\bea \Phi &=& {\tilde \rho} + k\w \alpha \nonumber \\ [.1in]
{*_\Phi} \Phi &=& {\hat {\tilde \rho}} \w \alpha + {\hat k}
\nonumber
\\ [.1in] {\sf B} &=& {\sf b} + {\sf a} \w \alpha \; . \eea
It will be convenient to write $e^{i\theta } =c+is$ and ${\tilde
\Omega} = {\tilde \rho} + i {\hat {\tilde \rho}}$ to define the new
3-forms $\rho = {\mbox{Re}} ( e^{-i\theta} {\tilde \Omega} )$, ${\hat \rho} = {\mbox{Im}} ( e^{-i\theta} {\tilde \Omega} )$. It
will also be convenient to take ${\hat k} = {1 \over 2} k\w k$,
anticipating the Calabi-Yau substructure that will occur in the
reduction. In terms of the new data, ({\ref{defrho}}) can be
written as
\be \varrho \; =\; -e^{-\varphi} \rho \w  e^{{\sf b} + {\sf a} \w
\alpha} + {\mbox{Re}} ( \, i e^{-\varphi-i\theta} e^{{\sf b}+ik} )
\w \alpha \; . \ee
Notice that the second term looks like it will give a stable
even-form spanning ${\mbox{exp}} ( \Lambda^0 \otimes {\mathbb{C}}
\oplus \Lambda^2 \otimes {\mathbb{C}} ) \subset \Lambda^{\rm even}
\otimes {\mathbb{C}}$ in six dimensions. The non-$\alpha$ terms
however seem to be missing a 1-form component needed in order to
reduce to a generic stable odd-form.

If we now reduce by restricting attention to special generalized
$G_2$ manifolds of the form $N \times S^1$, with generalized
Calabi-Yau structure on $N$ and coordinate $t \in S^1$, then one
can identify $\alpha = dt + \zeta$, where $\zeta$ is an arbitrary
harmonic 1-form on $N$ (this preserves the constraints that
$\alpha$ be closed and have unit norm with respect to the metric
reconstructed from $\Phi$). It is $\zeta$ that will account for the missing 1-form above. It
is perhaps worth noting that if one dropped the harmonic
constraint on $\zeta$, and only assumed it was closed, then one
could always reobtain a harmonic representative in the cohomology
class $[ \zeta ]$ via a suitable shift $\zeta \rightarrow \zeta +
d \gamma$ resulting from the 7-dimensional diffeomorphism
generated by the 7-vector $X$ whose only non-vanishing component
is $X^t = \gamma$ (i.e. a scalar on $N$). Similarly, one can use
the freedom to shift ${\sf B} \rightarrow {\sf B} + d \varepsilon$
in seven dimensions to remove the term ${\sf a}\w dt$ in the
reduced 2-form ${\sf B}$ by choosing $\varepsilon = t\,  {\sf a}$.
We will not do this however since we do not yet want to fix any of
the symmetries of the reduced theory.

The explicit expressions for the reduction of the stable forms in
the generalized $G_2$ theory are
\bea \varrho &=& e^{-\varphi} \Big[ s \zeta + \left\{ -\rho + (s{\sf b}-ck)\w \zeta \right\}  \nonumber \\
&&\qquad + \left. \left\{ -\rho\w {\sf b} + \left( {s\over 2} ({\sf b}^2 - k^2 ) - c{\sf b}\w k - \rho\w {\sf a} \right) \w \zeta \right\} \right] \nonumber \\ [.1in] &&+ e^{-\varphi} dt \w \left[ s + \left\{ s{\sf b}-ck \right\} + \left\{ {s\over 2} ({\sf b}^2 - k^2 ) - c{\sf b}\w k - \rho\w {\sf a} \right\} \right. \nonumber \\
[.1in] && \left. + \left\{ {s\over 6} ({\sf b}^3 - 3 k^2 \w {\sf b} ) + {c \over 6} ( k^3 - 3 {\sf b}^2 \w k ) - \rho\w {\sf b}\w {\sf a} \right\} \right] \eea
\bea
{\hat \varrho} &=& e^{-\varphi} dt \w \left[ c{\sf a} + \left\{ {\hat \rho} -(sk+c{\sf b})\w {\sf a} \right\} + \left\{ -{\hat \rho}\w {\sf b} - \left( {c\over 2} (k^2 - {\sf b}^2 ) - s{\sf b}\w k \right) \w {\sf a} \right\} \right] \nonumber \\ [.1in] &&+ e^{-\varphi} \Big[ c + \left\{ -sk -c{\sf b}-c {\sf a}\w \zeta \right\} \nonumber \\
&& + \left. \left\{ {c\over 2} ({\sf b}^2 - k^2 ) +s{\sf b}\w k - {\hat \rho}\w \zeta + (sk+c{\sf b})\w {\sf a}\w \zeta  \right\}  \right. \nonumber \\
[.1in] &&\left. + \left\{ {s\over 6} (k^3 - 3 {\sf b}^2 \w k ) - {c \over 6} ( {\sf b}^3 - 3 k^2 \w {\sf b} ) + \left( {c \over 2} ( k^2 - {\sf b}^2 ) -s{\sf b}\w k \right) \w {\sf a}\w \zeta \right. \right. \nonumber \\
&&\qquad + {\hat \rho}\w {\sf b}\w \zeta \Big\} \Big] \; . \label{genred1} \eea
This allows us to identify $\varrho = \chi_- + \chi_+ \w dt$ and ${\hat \varrho} = {\hat \chi_-} \w dt + {\hat \chi_+}$, in terms of odd-forms $\chi_-$, ${\hat \chi_-}$ and even-forms $\chi_+$, ${\hat \chi_+}$ on $N$. In fact, it will be more convenient to define $\varrho = e^{{\sf a} \w \zeta} \w \chi_- + \chi_+ \w dt$ and ${\hat \varrho} = {\hat \chi_-} \w dt + e^{-{\sf a} \w \zeta} \w {\hat \chi_+}$. The identities $\int_N \chi_- \w {\sf a} \w \zeta \w {\hat \chi_-} =0$ and $\int_N \chi_+ \w {\sf a} \w \zeta \w {\hat \chi_+} =0$ ensure that either choice will give rise to the same reduced Hitchin functional $\int_M \varrho \w {\hat \varrho} =\int_{N\times S^1} ( \chi_- \w {\hat \chi_-} + \chi_+ \w {\hat \chi_+} )\w dt$. Thus we have
\bea \chi_- &=& e^{-\varphi} e^{{\sf b}} \w \left[ s \zeta - (\rho + c\, \zeta \w k) - {1\over 2} k^2 \w s \zeta  \right] \nonumber \\ [.1in]
\chi_+ &=& e^{-\varphi} e^{{\sf b}} \w \left[ s -c k - ( \rho \w {\sf a} + {s\over 2} k^2 ) + {c\over 6} k^3 \right] \nonumber \\ [.1in]
{\hat \chi_-} &=& e^{-\varphi} e^{-{\sf b}} \w \left[ -c {\sf a} - ( {\hat \rho} - s {\sf a} \w k) + {1\over 2} k^2 \w c {\sf a}  \right] \nonumber \\ [.1in]
{\hat \chi_+} &=& e^{-\varphi} e^{-{\sf b}} \w \left[ c -s k - ( {\hat \rho} \w \zeta + {c\over 2} k^2 ) + {s\over 6} k^3 \right] \; . \label{genred1a} \eea
Notice that the odd-forms are related by an anti-involution $\chi_- \rightarrow {\hat \chi_-}$, ${\hat \chi_-} \rightarrow -\chi_-$ that is generated by the parameter transformations $(s,c,\rho , k, {\sf b}, {\sf a} , \zeta ) \rightarrow (c,-s,{\hat \rho} , -k, -{\sf b}, -\zeta , -{\sf a} )$. This is a symmetry of the odd-form functional $\int_{N\times S^1} \chi_- \w {\hat \chi_-} \w dt$ if $t$ is invariant. Similarly the even-forms are related by an anti-involution $\chi_+ \rightarrow {\hat \chi_+}$, ${\hat \chi_+} \rightarrow -\chi_+$ that is generated by $(s,c,\rho , k, {\sf b}, {\sf a} , \zeta ) \rightarrow (c,-s,{\hat \rho} , -k, -{\sf b}, \zeta , {\sf a} )$. This is a symmetry of the even-form functional \\  $\int_{N\times S^1} \chi_+ \w {\hat \chi_+} \w dt$ if $t \rightarrow -t$. In both cases the transformations of $(s,c, \rho )$ follow from a shift $\theta \rightarrow \theta + \pi /2$ of the angle between the two $G_2$-invariant unit spinors in 7 dimensions (recalling that $\rho = c{\tilde \rho} + s {\hat{\tilde \rho}}$ and ${\hat \rho} = -s{\tilde \rho} + c {\hat{\tilde \rho}}$ in terms of the original data).

These two anti-involutions just correspond to the action of the Hamiltonian vector field on the symplectic spaces spanned by $\chi_\pm + i {\hat \chi_\pm}$ that is defined respectively by the odd- and even-form functionals
\bea \int_N \chi_- \w {\hat \chi_-} &=& \int_N e^{-2\varphi} \left[ \rho \w {\hat \rho} - s \rho \w k \w {\sf a} - c {\hat \rho} \w k \w \zeta  \right] \nonumber \\ [.1in]
\int_N \chi_+ \w {\hat \chi_+} &=& \int_N e^{-2\varphi} \left[ {2\over 3} k^3 + s \rho \w k \w {\sf a} + c {\hat \rho} \w k \w \zeta  \right] \; , \label{genodev} \eea
using the constant symplectic form on the complexified stable odd- and even-form spaces, as described by Hitchin on p.16 in {\cite{NH3}}. Thus the circle action $\chi_\pm + i {\hat \chi_\pm} \rightarrow e^{-i \vartheta} ( \chi_\pm + i {\hat \chi_\pm} )$ generated by this Hamiltonian vector field has a nice interpretation via shifts $\theta \rightarrow \theta + \vartheta$ in the angular separation of the generalized $G_2$ unit spinors.

As a quick consistency check, we see that the sum of these functionals
\be
\int_M \varrho \w {\hat \varrho} \; =\; \int_{N\times S^1} e^{-2\varphi} \left( \rho\w {\hat \rho} + {2 \over 3} k\w k\w k \right)\w dt \; \label{genred2} ,
\ee
corresponds to the generalized Calabi-Yau Hitchin functional {\cite{GLW}}, modulo the compatibility conditions that we will now discuss.

With the identification ({\ref{genred1a}}), the generalized Calabi-Yau compatibility conditions ({\ref{gencom}}) for $\chi_\pm$ become
\bea
c \left( \rho + {1 \over c} \zeta \w k \right) \w \left( k + {s \over c} {\sf a}\w \zeta \right) &=& 0 \nonumber \\ [.1in]
\left[ ( \rho + c\, \zeta\w k ) \w {\sf a} + {s \over 2} k^2 \right] \w \iota_v \rho &=& 0 \; , \label{gencom2}
\eea
for any vector field $v$ (the arbitrary 1-form $\xi$ has been factored out of the first equation). An additional term $\left[ c \rho\w k +s \rho \w {\sf a} \w \zeta + k^2 \w \zeta \right] \w \iota_v {\sf b}$ in the second equation vanishes as a result of the first equation, to completely remove the dependence on ${\sf b}$ in ({\ref{gencom2}}). The second equation has also been simplified using $c \rho\w k\w \zeta =0$ which follows from the first equation.

It will be useful to conclude this subsection by also noting the related compatibility conditions for ${\hat \chi_\pm}$
\bea
s \left( {\hat \rho} - {1 \over s} {\sf a} \w k \right) \w \left( k + {c \over s} {\sf a}\w \zeta \right) &=& 0 \nonumber \\ [.1in]
\left[ ( - {\hat \rho} + s\, {\sf a} \w k ) \w \zeta + {c \over 2} k^2 \right] \w \iota_v {\hat \rho} &=& 0 \; . \label{gencom2a}
\eea
%


\subsubsection{Comparison with Pestun-Witten}

Let us now consider how the quadratic part of the reduced
functional above relates to the one considered in {\cite{PW}}.
Recall that the extended functional used in {\cite{PW}}
corresponds to the Hitchin functional for a stable odd-form
$\sigma$, fixed by $SU(3,3)$. Since the reduction of the
generalized $G_2$ Hitchin functional involves both a stable
odd-form $\chi_-$ and even-form $\chi_+$, fixed by $SU(3)\times
SU(3) \subset G_2 \times G_2$, we should not expect these theories
to agree directly but will find that the theory of Pestun and
Witten arises as a truncation of the generalized $G_2$ theory,
with $\chi_-$ related to the odd-form $\sigma$ in {\cite{PW}}
after imposing the generalized Calabi-Yau compatibility equations.

When expanded around a Calabi-Yau background $N_0$, the quadratic part of the extended functional in {\cite{PW}} becomes
\be
\int_{N_0} \delta \sigma \w \delta {\hat \sigma} \; =\; \int_{N_0} \delta \sigma^3 \w J \delta \sigma^3 + 2 \delta \sigma^5 \w J \delta \sigma^1 \; ,
\ee
where $\delta {\hat \sigma} = J \delta \sigma$ in terms of the background complex structure $J : \Lambda^{\rm odd} \rightarrow \Lambda^{\rm odd}$. The action of this background complex structure can be written $J = J_1 + J_3 + J_5$ where $J_3$ acts on 3-forms via the map given below ({\ref{6quad}}), $J_1$ acts on 1-forms mapping $\xi_m \rightarrow k_m^{\;\;\; n} \xi_n$ and $J_5 = - * J_1 *$ acting on 5-forms
\footnote{The sign here follows from the requirement that $J_5 \omega \w J_1 \xi = \omega \w \xi$ for any 5-form $\omega$ and 1-form $\xi$. In particular, taking $\omega = *\xi$ then this follows from the fact that $* J_1 \xi \w J_1 \xi = *\xi \w \xi$ and that $*^2 =-1$ on odd-forms in 6 dimensions.}
. Thus one has $J^2 =-1$. The identity $\sigma^1 \w J \delta \sigma^5 = \sigma^5 \w J \delta \sigma^1$ has been used above.

To relate this to the quadratic part of the odd-form functional for $\chi_-$ in ({\ref{genodev}}) requires some more work. We begin by implementing the first compatibility equation in ({\ref{gencom2}}), ({\ref{gencom2a}}) in ({\ref{genodev}}) which gives
\bea \int_N \chi_- \w {\hat \chi_-} &=& \int_N e^{-2\varphi} \left[ \rho \w {\hat \rho} - {1 \over sc} k^2 \w {\sf a}\w \zeta  \right] \nonumber \\
 &=& \int_N e^{-2\varphi} \left( \rho + {1 \over c} \zeta \w k \right) \w \left( {\hat \rho} - {1 \over s} {\sf a} \w k  \right)  \nonumber \eea
\bea
\int_N \chi_+ \w {\hat \chi_+} &=& \int_N e^{-2\varphi} \left[ {2\over 3} k^3 + {1 \over sc} k^2 \w {\sf a}\w \zeta  \right] \nonumber \\
 &=& \int_N e^{-2\varphi} k \w \left( k + {s \over c} {\sf a} \w \zeta \right) \w \left( k + {c \over s} {\sf a} \w \zeta \right)  \; , \nonumber \\ \eea
for generic $s, c \neq 0$. Notice this has decoupled the terms involving $\rho$ and ${\hat \rho}$ from ${\sf a}$ and $\zeta$ in the odd-form functional.

Although we have not yet expanded around a fixed Calabi-Yau background it will be convenient to define the map $J_1 : \Lambda^1 \rightarrow \Lambda^1$ as $J_1 ( \xi ) = -* \left( {1 \over 2} k^2 \w \xi \right)$ and the map $J_5 = -* J_1 * : \Lambda^5 \rightarrow \Lambda^5$ which will reduce to their namesakes defined earlier in the quadratic expansion. The odd-form functional above can then be written more suggestively as
\be
\int_{N} \chi_- \w {\hat \chi_-} \; =\; \int_{N} e^{-2\varphi} \left[ \rho \w {\hat \rho} + {2 \over sc} {*{\sf a}} \w J_1  \zeta   \right] \; .
\ee
From this one can read off more suitable expressions for the odd-forms
\bea \chi_- &=& e^{-\varphi} e^{{\sf b}} \w \left[ {1 \over c} \zeta + \rho + {1\over s} {*{\sf a}}  \right] \nonumber \\ [.1in]
{\hat \chi_-} &=& e^{-\varphi} e^{-{\sf b}} \w \left[ {1 \over c} J_1 \zeta + {\hat \rho} + {1\over s} {*J_1 {\sf a}} \right]  \; , \label{genred1b} \eea
which are presumably related to those given in ({\ref{genred1a}}) by a suitable symplectic transformation on the space of stable odd-forms (supplemented with the generalized Calabi-Yau compatibility constraints), since they both give rise to the same Hitchin functional
\footnote{For example, the expressions for $\chi_-^5$ and ${\hat \chi_-}^1$ inside the square brackets in ({\ref{genred1a}}) and ({\ref{genred1b}}) are related by the symplectic transformation $\chi_-^5 \rightarrow -*{\hat \chi_-}^1$, ${\hat \chi_-}^1 \rightarrow *\chi_-^5$ that preserves $\int_N \chi_-^5 \w {\hat \chi_-}^1$ (followed by a multiplicative factor $1/sc$ related to the generalized Calabi-Yau constraint).}
. One therefore has $\delta {\hat \chi_-} = J \delta \chi_-$ for first order variations around a Calabi-Yau background with K\"{a}hler form $k$ and complex structure $\Omega = \rho + i {\hat \rho}$. One can recover the quadratic functional of Pestun and Witten by identifying the first order variations $\delta \sigma^1 = \delta ( c^{-1} \zeta )$, $\delta \sigma^3 = \delta ( e^{-\varphi} \rho )$ and $\delta \sigma^5 = * \delta ( s^{-1} {{\sf a}} )$.


\section{Background dependence} \label{backdep}
\setcounter{equation}{0}

In this section we will investigate the dependence of the 1-loop
partition functions we have calculated on the choice of background
metric. This can be deduced from theorems in {\cite{SC}} but we
will derive it from first principles. This analysis will help us reconcile the results of sections 5 and 6, by showing how the 1-loop partition functions $Z_{gen}$ and $Z_{string}$ are related. The pertinent quantity to
calculate is the first order variation $\delta_g ( {\mbox{log
det}}\, \Delta^p )$ of the logarithm determinant of Laplacians
acting on $p$-forms in $D$ dimensions. For simplicity we will
begin by assuming metric variations around a background with
trivial cohomology so that there are no extra contributions from
harmonic forms to concern us.


\subsection{General formulae}

Consider the variation of the canonical inner product
\be \langle \omega , \xi \rangle_p \; =\; \int_M d^D x \sqrt{g} \,
\frac{1}{p!} \, g^{m_1 n_1} ... g^{m_p n_p} \, \omega_{m_1 ...
m_p} \xi_{n_1 ... n_p} \; , \label{innprod} \ee
between two $p$-forms on a $D$-dimensional Riemann manifold $M$, with
respect to the Riemannian metric $g$. This can be written
\be \delta_g \langle \omega , \xi \rangle_p \; =\; \langle \omega
, B_p \, \xi \rangle_p \; =\; \langle B_p \, \omega , \xi
\rangle_p \; ,\ee
in terms of the algebraic operator
\be ( B_p )_{m_1 ... m_p}^{\quad\quad\;\;\;\, n_1 ... n_p} \; =\;
p \, \delta g^{ab} \delta_b^{[n_1} g_{a [m_1} \delta_{m_2}^{n_2}
... \delta_{m_p ]}^{n_p ]} - \frac{1}{2} \delta g^{ab}
\delta_{[m_1}^{n_1} ... \delta_{m_p ]}^{n_p} \; , \ee
which is a function of $\delta g$ and $g$, mapping $\Lambda^p
\rightarrow \Lambda^p$.

One can prove that $\delta_g (*\omega ) = - B_{D-p} {*\omega} =
{*B_p} \, \omega$ for any $\omega \in \Lambda^p$, from which one
derives
\be \delta_g ( d^\dagger \omega ) = d^\dagger B_p \omega - B_{p-1}
d^\dagger \omega \; . \ee

Using the formula $\delta ( {\mbox{log det}} X) = \delta (
{\mbox{det}} X) / | {\mbox{det}} X | = {\mbox{tr}} ( X^{-1} \delta
X )$, for the variation of an elliptic operator $X$, and $\Delta^p
= d^\dagger_{p+1} d_p + d_{p-1} d^\dagger_p$, one obtains the
result
\be \label{metvar} \delta_g ( {\mbox{log det}}\, \Delta^p ) \; =\; -2\,
{\mbox{tr}} \left( B_p + 2 \sum_{k=0}^{p-1} B_k \right) \; =\;  - {D
\choose p} {\mbox{tr}}\left( g^{-1} \delta g \right) \; . \ee
Notice that ${\mbox{tr}}\left( g^{-1} \delta g \right) = 2\, \delta_g ( {\mbox{log Vol}}(M) )$, where ${\mbox{Vol}}(M)$ is the volume of the Riemann manifold $M$. Some partition functions that are $\delta_g$-invariant around backgrounds with trivial cohomology develop a gravitational anomaly for variations around more general backgrounds. Topological symmetry can sometimes be restored in such cases by multiplying the original partition function by a compensating power of the volume of the background manifold. This has been shown for the B-model in {\cite{PW}} (and previously in unpublished work by Klemm and Vafa) and for certain Chern-Simons type actions in {\cite{SC}}.

A nice consistency check of this result is to verify the lemma of
Schwarz {\cite{SC}} which states that the Ray-Singer torsion is a
topological invariant in odd dimensions. Taking the log of this
torsion, for $D=2k+1$, one finds that indeed
\be \delta_g \left( -\frac{1}{2} \sum_{p=0}^k (-1)^p (2p+1)
{\mbox{log det}}\, \Delta^{k-p} \right) \; =\; {\mbox{tr}} \left(
\sum_{p=0}^{k} (-1)^{k-p} B_p \right) \; =\; 0 \; .  \ee
The last equality simply follows from a combinatorial identity.


\subsection{Metric dependence of 1-loop Hitchin functionals}

Using the expressions found in the last subsection, one can
calculate the metric variation of the (log of the) 1-loop $G_2$
partition functions $Z$ ({\ref{partf2}}), $Z_{gen}$
({\ref{genpart}}), $Z_{string}$ ({\ref{stringpart}}). None of them are $\delta_g$-invariant. This is to be expected though since the $G_2$ metric deformations contain both K\"{a}hler and complex structure deformations in the reduced theory. $\delta_g$-invariance in the B-model is related to it not depending on K\"{a}hler moduli, but it has the well-known wavefunction behaviour under variations of the complex structure of the Calabi-Yau.

To understand this in more detail, let us examine the structure of
topological invariants built out of products of powers of
${\mbox{det}}\, \Delta^p$ in $D=7$. Since Hodge duality implies
${\mbox{det}}\, \Delta^p = {\mbox{det}}\, \Delta^{D-p}$, the only
independent Laplacian determinants are for $p=0,1,2,3$. (Using the
various $G_2$ isomorphisms already mentioned, all the 1-loop
partition functions we have calculated can be written as products
of powers of these 4 Laplacian determinants.) Consider now the
most general such product
$$
({\mbox{det}}\, \Delta^3 )^a ({\mbox{det}}\, \Delta^2 )^b
({\mbox{det}}\, \Delta^1 )^c ({\mbox{det}}\, \Delta^0 )^d \; ,
$$
specified by any 4 real numbers $a$, $b$, $c$, $d$. Demanding the
log of this expression to be $\delta_g$-invariant implies $35a
+21b +7c+d=0$. The 3 independent numbers parameterizing the
invariant can be recast as powers of 3 more basic topological
invariants. A convenient choice for these 3 basis invariants is
\bea {\sf I}_0 &=& ({\mbox{det}}\, \Delta^1 )^{-1/2} ({\mbox{det}}\, \Delta^0 )^{7/2} \; =\; ({\mbox{det}}\, \Delta_{\bf 7} )^{-1/2} ({\mbox{det}}\, \Delta_{\bf 1} )^{7/2} \nonumber \\
[.1in] {\sf I}_1 &=& ({\mbox{det}}\, \Delta^2 )^{-1/2} ({\mbox{det}}\, \Delta^1 )^{3/2} \; =\; ({\mbox{det}}\, \Delta_{\bf 14} )^{-1/2} ({\mbox{det}}\, \Delta_{\bf 7} ) \nonumber \\
[.1in] {\sf I}_2 &=& I_{RS} \; =\; ({\mbox{det}}\, \Delta^3
)^{-1/2} ({\mbox{det}}\, \Delta^2 )^{3/2} ({\mbox{det}}\, \Delta^1
)^{-5/2} ({\mbox{det}}\, \Delta^0 )^{7/2} \nonumber \\ [.1in] &=&
({\mbox{det}}\, \Delta_{\bf 27} )^{-1/2} ({\mbox{det}}\,
\Delta_{\bf 14} )^{3/2} ({\mbox{det}}\, \Delta_{\bf 7} )^{-3/2}
({\mbox{det}}\, \Delta_{\bf 1} )^3
 \; . \label{3top7} \eea
Any $\delta_g$-invariant constructed from products of powers of
Laplacian determinants in $D=7$ can be written as ${\sf I}_0^a
{\sf I}_1^b {\sf I}_2^c$, for some choice of $a$, $b$, $c$. $Z$,
$Z_{gen}$ and $Z_{string}$ cannot be factorized in this way.
However, $Z$, $Z_{gen}$ and $Z_{string}$ can be written
\bea Z &=& {\sf I}_1 {\sf I}_0^{-1} \times {\mbox{Tor}} ( {\check
D}) \times ({\mbox{det}}\, \Delta_{\bf 1} )^{1/2} \nonumber \\ [.1in]
Z_{gen} &=& {\sf I}_2^{-2} {\sf I}_1 {\sf I}_0^{-1} \times
{\mbox{Tor}} ( {\check D}) \nonumber
\\ [.1in]
Z_{string} &=& {\sf I}_2 {\sf I}_1^4 {\sf I}_0^{-4} \times
{\mbox{Tor}} ( {\check D})^4 \; , \label{g2fact} \eea
in terms of the invariants ${\sf I}_0$, ${\sf I}_1$, ${\sf I}_2$,
and a non-invariant object ${\mbox{Tor}} ( {\check D}) =
({\mbox{det}}\, \Delta_{\bf 7} )^{-1/2} ({\mbox{det}}\,
\Delta_{\bf 1} )^{3/2}$ corresponding to the analytic torsion of
the $G_2$ Dolbeaux ${\check D}$ complex in ({\ref{g2dol}}). Recall that the ${\check D}$ complex describes the spectrum of the topological $G_2$ string {\cite{g2string}} in much the same way that the ${\bar \partial}$ complex does for the B-model. Indeed $Z_{string} = {\mbox{Tor}} ( {\check D} \otimes \Lambda^1 ) / {\mbox{Tor}} ( {\check D} )^3$, where ${\mbox{Tor}} ( {\check D} \otimes \Lambda^1 ) = ({\mbox{det}}\, \Delta_{{\bf 7} \otimes {\bf 7}} )^{-1/2} ({\mbox{det}}\,
\Delta_{\bf 7} )^{3/2} = ({\mbox{det}}\, \Delta_{\bf 27} )^{-1/2}
({\mbox{det}}\, \Delta_{\bf 14} )^{-1/2} ( {\mbox{det}}\,
\Delta_{\bf 7} ) ({\mbox{det}}\, \Delta_{\bf 1} )^{-1/2}$ is the analytic torsion of the ${\check D}$ complex for $\Lambda^1_{\bf 7}$-valued forms. $Z_{string}$ is therefore not $\delta_g$-invariant due to the identity ${\mbox{Tor}} ( {\check D} \otimes \Lambda^1 ) = {\sf I}_2 {\sf I}_1^4 {\sf I}_0^{-4} \times {\mbox{Tor}} ( {\check D})^7$.

The expressions above show that $Z_{gen} = I_{RS}^{-9/4}\, Z_{string}^{1/4}$. Therefore, although not identical, the 1-loop partition functions for the generalized $G_2$ Hitchin functional and the topological $G_2$ string seem to be related up to a power of the Ray-Singer torsion invariant of the background $G_2$ manifold.

Let us now perform a similar analysis in $D=6$ on a Calabi-Yau
manifold, to reconcile the results above with those found in
{\cite{PW}}. Hodge duality in $D=6$ again implies the only
independent Laplacian determinants are for $p=0,1,2,3$. However,
since we are considering a Calabi-Yau background, we can use the
Hodge decomposition $\Lambda^3 = \Lambda^{30} \oplus \Lambda^{21}
\oplus \Lambda^{12} \oplus \Lambda^{03}$ and vector space
isomorphisms $\Lambda^{30} = \Lambda^{03} = \Lambda^{00}$,
$\Lambda^{21} = \Lambda^{12} = \Lambda^{11}$ to relate 3-form
Laplacian determinants to ones for forms of lower degree. Thus the
most general product is of the form
$$
({\mbox{det}}\, \Delta^2 )^a ({\mbox{det}}\, \Delta^1 )^b
({\mbox{det}}\, \Delta^0 )^c \; .
$$
This is $\delta_g$-invariant if $15a+6b+c=0$. A convenient choice
of 2 basis invariants, whose powers are parameterized by these two
independent numbers, is
\bea I_0 &=& ({\mbox{det}}\, \Delta^1 )^{-1/4} ({\mbox{det}}\, \Delta^0 )^{3/2} \; =\; ({\mbox{det}}\, \Delta^{10} )^{-1/2} ({\mbox{det}}\, \Delta^{00} )^{3/2} \nonumber \\
[.1in] I_1 &=& ({\mbox{det}}\, \Delta^2 )^{-1/2} ({\mbox{det}}\,
\Delta^1 )^{5/4} \; =\; ({\mbox{det}}\, \Delta^{11} )^{-1/2}
({\mbox{det}}\, \Delta^{10} )^{3/2} \; . \label{2top6} \eea
These are precisely the holomorphic Ray-Singer torsions $I_0 =
I^{RS}_{0\, {\bar \partial}}$ and $I_1 = I^{RS}_{1 \, {\bar
\partial}}$ of the Dolbeault complex (for $\Lambda^{00}$- and
$\Lambda^{10}$-valued $(0,q)$-forms), used in {\cite{PW}}. The
1-loop partition function for the stable 3-form Hitchin functional
in $D=6$ is $I_1/I_0$ while that for the extended Hitchin
functional (and B-model) is $I_1/I_0^3$. Thus we see that both are
topological invariants
\footnote{If we drop the assumption of trivial cohomology of the
background, then the B-model partition function gets a
gravitational anomaly {\cite{PW}} and it must be multiplied by a
compensating volume factor ${\mbox{Vol}} (CY_3 )^{-\chi /12}$
(where $\chi$ is the Euler number of the Calabi-Yau background) to
make it invariant.}
.


\subsection{B-model gravitational anomaly from 7 dimensions?}

Notice that the coefficient in ({\ref{metvar}}) corresponds to the dimension of the vector space $\Lambda^p$. For backgrounds with non-trivial cohomology, ${\mbox{det}}\, \Delta^p$ is defined by removing the zero-modes of $\Delta^p$, corresponding to harmonic $p$-forms, from the determinant. Let us naively assume this results in the redefined coefficient ${D \choose p} \rightarrow {D \choose p} - b_p$ in ({\ref{metvar}}), where $b_p = {\mbox{dim}} ( H_p (M, {\mathbb{R}}) )$ are the Betti numbers. This will imply that partition functions that are invariant under deformations of a background metric with trivial cohomology can develop gravitational anomalies for variations around more general backgrounds. Furthermore, such anomalies will be proportional to some power of the volume of the background manifold. This power being some linear sum of Betti numbers.

For the B-model, the particular linear sum of Betti numbers appearing corresponds to the Euler number $\chi = 2( h_{11} - h_{12} )$, in terms of the Hodge numbers of the Calabi-Yau background. Let us now consider a $G_2$ background of the form $CY_3 \times S^1$ and ask if there is any combination of Laplacian determinants in 7 dimensions, whose metric variation correctly reduces to this B-model gravitational anomaly? Using the Kunneth formula $b_p ( CY_3 \times S^1 ) = b_p ( CY_3 ) + b_{p-1} (CY_3 )$, we see that $b_3 = 2(1+ h_{12}) + h_{11}$, $b_2 = h_{11}$ and $b_1 = b_0 =1$ on $CY_3 \times S^1$. Thus we can write
$$
\chi \; =\; 2( h_{11} - h_{12} ) \; =\; -b_3 + 3\, b_2 + (2-a) \, b_1 + a\, b_0 \; ,
$$
for any real number $a$. This coefficient appears from the metric variation of
$$
({\mbox{det}}\, \Delta^3 )^{-1/2} ({\mbox{det}}\, \Delta^2 )^{3/2} ({\mbox{det}}\, \Delta^1 )^{1-a/2} ({\mbox{det}}\, \Delta^0 )^{a/2} \; =\; I_{RS} \times \left( \frac{{\mbox{det}}\, \Delta^1}{{\mbox{det}}\, \Delta^0} \right)^{(7-a)/2} \; ,
$$
(possibly to some overall power). Of course, we also want this to be $\delta_g$-invariant around backgrounds with trivial cohomology and so must choose $a=7$. Hence the gravitational anomaly of the B-model on $CY_3$ has a simple interpretation from metric variations of the Ray-Singer torsion on $CY_3 \times S^1$.


\section{Conclusions and open questions} \label{concl}
\setcounter{equation}{0}

In this paper we have attempted to understand more about the quantum structure of topological M-theory {\cite{M}} from perturbative quantization of the (generalized) $G_2$ Hitchin functional.

We computed the 1-loop partition function of the ordinary $G_2$ Hitchin functional and agreement was found between the local degrees of freedom for the reduction of this theory on a circle and the corresponding theory of Pestun and Witten {\cite{PW}}, obtained from the Hitchin functional for a stable 3-form in 6 dimensions.

The calculation was repeated for the generalized $G_2$ Hitchin functional and a certain truncation of the circle reduction of this theory was related to the extended Hitchin functional in 6 dimensions, whose 1-loop partition function was equated with the topological B-model in {\cite{PW}}. The 1-loop partition function for the topological $G_2$ string {\cite{g2string}} was also computed here and found to agree with the generalized $G_2$ theory only up to a power of the Ray-Singer torsion of the background $G_2$ manifold.

There are however a number of subtleties involved in this
calculation. First it is not obvious to us that the linear
variations $\delta \varrho$ of the stable odd-form in 7 dimensions
constitute the appropriate degrees of freedom describing the
quantum theory. That is $\varrho$ is a non-linear function of
parameters which seem more naturally related to stringy moduli. To
clarify this point as well as for physical applications of our
results it would be important to understand whether our
computation could be related to effective actions for generalized
$G_2$ compactifications of physical string and M-theory. This
could also help determine the fundamental degrees of freedom of
topological M-theory.

Another important issue is whether the gauge field components we have ignored in the quantization, because they do not appear in the quadratic action (i.e. they are projected out or neglected as total derivatives), have a non-trivial contribution to the partition function. For instance they could be important in defining an appropriate path integral measure and give rise to non-trivial 1-loop determinants that would modify our results. It is possible that resolving these subtleties could lead to a more precise agreement between the generalized $G_2$ Hitchin functional and topological $G_2$ string at 1-loop.

Certainly it would be interesting to understand the global structure of the 1-loop $G_2$ Hitchin functionals for topologically non-trivial gauge fields and their reduction to 6 dimensions. According to the philosophy of topological M-theory, this would provide a non-trivial gauge-theoretic description of the coupling between the Pestun-Witten description of the B-model {\cite{PW}} and the quantum foam description of the A-model {\cite{qf1,qf2}}. The observables of this theory could compute interesting gerbe invariants.

Higher order diffeomorphism-invariant terms in the expansion of the $G_2$ Hitchin functional can be understood as BRST-invariant operators deforming the quadratic theory. For example, in the reduction $B = b +A\w dt$, the cubic term in the expansion that we calculated in section 2.2 contains the $\int F \w F \w F$ deformation of the quadratic term $\int k \w F \w F$ in the A-model quantum foam {\cite{qf1,qf2}}. It would be interesting to understand the effect of such higher order deformations in 7 dimensions.

Finally, since general background $G_2$ metric variations contain complex structure variations in 6 dimensions, it is natural to ask whether the wavefunction behaviour of B-model has a nice interpretation in 7 dimensions? Indeed this was one of the original motivations for the proposal of topological M-theory in {\cite{M}}. It is possible that this could be understood from the structure of partition functions we have calculated here although we have not investigated this idea.


\section*{\large{Acknowledgments}}

We would like to thank Jos\'{e} Figueroa-O'Farrill, Lotte
Hollands, Andy Neitzke, Martin Ro\v{c}ek and Cumrun Vafa for
helpful discussions related to this work. In particular, we also
thank David Berman for sharing with us an earlier attempt to
compute the one-loop partition function of the 7d Hitchin
functional, and Asad Naqvi for help in determining the one-loop
partition function of the topological $G_2$ string. The work of
PdM is supported in part by a Seggie Brown fellowship. The work of
AS is supported by the EC Marie Curie Research Training Grant
MRTN-CT-2004-512194 of the Superstrings Network. The work of JdB
and SES is supported financially by the Foundation of Fundamental
Research on Matter (FOM).


\appendix

\section{\large{$G_2$ and CY identities}}
\setcounter{equation}{0}

A seven-dimensional Riemann manifold $M$ is guaranteed to have holonomy in the subgroup of $G_2 \subset SO(7)$ by the existence of a harmonic 3-form $\Phi$.
In our conventions $\Phi$ and its Hodge-dual $*\Phi$ can be written
\bea
\Phi &=& e^{123} - e^{147} - e^{156} - e^{246} + e^{257} + e^{345} + e^{367} \nonumber \\
*\Phi &=& e^{1245} + e^{1267} + e^{1346} - e^{1357} - e^{2347} - e^{2356} + e^{4567} \label{orth} \; ,
\eea
with respect to an orthonormal basis $e^I$ (where $e^{I_1 ... I_p} = \,  e^{I_1} \wedge ... \wedge e^{I_p}$).

Some useful identities for products of the components of $\Phi$ and $*\Phi$ are as follows
\bea
{*\Phi}_{IJKA} {*\Phi}^{PQRA} &=& 6\, \delta_{[I}^P \delta_J^Q \delta_{K]}^R +9\, \delta_{[I}^{[P} {*\Phi}_{JK]}^{\;\;\;\;\; QR]} - \Phi_{IJK} \Phi^{PQR} \nonumber \\
{*\Phi}_{IJKA} \Phi^{PQA} &=& -6\, \delta_{[I}^{[P} \Phi_{JK]}^{\;\;\;\;\;\, Q]} \nonumber \\
\Phi_{IJA} \Phi^{PQA} &=& 2\, \delta_{[I}^P \delta_{J]}^Q + {*\Phi}_{IJ}^{\;\;\;\; PQ} \label{iden} \; .
\eea
These identities can be proven in the orthonormal basis above but it is clear they are also valid in any coordinate basis by simply
acting on the formulae with the appropriate combination of vielbeins.
Other required identities follow by taking contractions or Hodge dualizations of the ones above.
For example,
\bea
{*\Phi}_{IJAB} {*\Phi}^{PQAB} &=& 8\, \delta_{[I}^P \delta_{J]}^Q + 2\, {*\Phi}_{IJ}^{\;\;\;\; PQ} \nonumber \\
{*\Phi}_{IJAB} \Phi^{PAB} &=& 4\, \Phi_{IJ}^{\;\;\;\;\, P} \nonumber \\
\Phi_{IJK} \, \epsilon^{ABCDEJK} &=& 10\, {*\Phi}^{[ABCD} \delta^{E]}_{I} \; .
\eea

One can deduce the corresponding Calabi-Yau identities from
dimensional reduction of the $G_2$ ones above, with $\Phi = \rho +
k\w dt$ and ${*\Phi} = {\hat \rho} \w dt + {1 \over 2} k\w k$
($\rho$ and ${\hat \rho}$ being the real and imaginary parts of
the holomorphic (3,0)-form $\Omega$ and $k$ being the K\"{a}hler
form), as in (\ref{bakred}). In an orthonormal basis $e^m$ for the
Calabi-Yau we can write $\Omega = -i\, dz^1 \w dz^2 \w dz^3$ and
$k = {i \over 2} dz^m \w d{\bar z}^m$, where $dz^m = e^m + i\,
e^{m+3}$. Written out explicitly, these expressions give
\bea
\rho &=& e^{126} - e^{135} + e^{234} -e^{456} \nonumber \\
{\hat \rho} &=& -e^{123} + e^{156} - e^{246} + e^{345} \nonumber \\
k &=& e^{14} + e^{25} + e^{36} \; , \eea
and follow from the aforementioned reduction of ({\ref{orth}})
after relabelling $1 \leftrightarrow 4$.

Some useful identities for products of components of $\rho$,
${\hat \rho}$ and $k$ are
\bea k_{mn} k^{pq} + \rho_{mnr} \rho^{pqr} &=& 2\, \delta_{[m}^p
\delta_{n]}^q + {1 \over 2} (k\w k)_{mn}^{\quad\, pq} \nonumber \\
\rho_{mnr} \rho^{pqr} &=& {\hat \rho}_{mnr} {\hat \rho}^{pqr} \; =
\; 2\, \delta_{[m}^p
\delta_{n]}^q - 2\, k_{[m}^{\;\;\, p} k_{n]}^{\;\;\, q} \nonumber \\
\rho_{mnr} {\hat \rho}^{pqr} &=& 2\, \delta_{[m}^p k_{n]}^{\;\;\,
q} - 2\, \delta_{[m}^q k_{n]}^{\;\;\, p} \nonumber \\
\rho_{mnp} \rho^{qrs} + {\hat \rho}_{mnp} {\hat \rho}^{qrs} &=&
6\, \delta_{[m}^q \delta_{n}^r \delta_{p]}^s - 18\, \delta_{[m}^q
k_{n}^{\;\;\, r} k_{p]}^{\;\;\, s} \nonumber \\ [.1in]
k_{mn} k^{np} &=& - \delta^p_m \nonumber \\
k_{mn} \rho^{npq} &=& - \hat{\rho}_m^{\;\;\; pq} \nonumber \\
k_{mn} \hat{\rho}^{npq} &=&  \rho_m^{\;\;\; pq} \nonumber \\
\rho_{mnp} \rho^{npq} &=& {\hat \rho}_{mnp} {\hat \rho}^{npq} \; =\; 4\, \delta_m^q \nonumber \\
\rho_{mnp} \hat{\rho}^{npq} &=& 4\, k_m^{\;\;\; q} \nonumber \\
[.1in]
\epsilon^{mnpqrs} \rho_{qrs} &=& -6 \, \hat{\rho}^{mnp} \nonumber \\
\epsilon^{mnpqrs}\hat{\rho}_{qrs} &=& 6\, \rho^{mnp} \nonumber \\
\epsilon^{mnpqrs} k_{rs} &=& 6\, k^{[mn} k^{p]q} \nonumber \\
[.1in] {1 \over 4} \rho \w \hat{\rho} &=& {1 \over 6} k\w k\w k \; =\; *1 \nonumber \\
5\, \rho_{[mnp} \hat{\rho}_{qrs]} &=& 15\, k_{[mn} k_{pq} k_{rs]}
\; =\; \epsilon_{mnpqrs} \; .\eea
%


\section{\large{$G_2$ cohomology}} \label{sec-gc}
\setcounter{equation}{0}

The de Rham cohomology groups on a seven-manifold $M$ with holonomy in $G_2$ have the following decompositions
\bea
H^0 (M,{\mathbb{R}}) &=& {\mathbb{R}} \nonumber \\
H^1 (M,{\mathbb{R}}) &=& H^1_{\bf{7}} (M,{\mathbb{R}}) \nonumber \\
H^2 (M,{\mathbb{R}}) &=& H^2_{\bf{7}} (M,{\mathbb{R}}) \oplus H^2_{\bf{14}} (M,{\mathbb{R}}) \nonumber \\
H^3 (M,{\mathbb{R}}) &=& H^3_{\bf{1}} (M,{\mathbb{R}}) \oplus H^3_{\bf{7}} (M,{\mathbb{R}}) \oplus H^3_{\bf{27}} (M,{\mathbb{R}}) \; .
\eea
Similar decompositions follow for the remaining cohomology groups by Hodge duality.
The subscripts in $H^I_{\bf{n}}$ denote the irreducible representations ${\bf{n}}$ of $G_2$ that the $I$-form components occupy.
The non-trivial projection operators ${\sf{P}}^I_{\bf{n}}$ onto these irreducible subspaces are given by
\bea
( {\sf{P}}^2_{\bf{7}} )_{IJ}^{\;\;\;\, PQ} &=& \frac{1}{6} \, \Phi_{IJA} \Phi^{PQA} \; =\; \frac{1}{3} \, \left( \delta_{[I}^P \delta_{J]}^{Q} + \frac{1}{2} \, {*\Phi}_{IJ}^{\;\;\;\, PQ} \right) \nonumber \\
( {\sf{P}}^2_{\bf{14}} )_{IJ}^{\;\;\;\, PQ} &=& \delta_{[I}^P \delta_{J]}^{Q} -\frac{1}{6} \, \Phi_{IJA} \Phi^{PQA} \; =\; \frac{2}{3} \, \left( \delta_{[I}^P \delta_{J]}^{Q} - \frac{1}{4} \, {*\Phi}_{IJ}^{\;\;\;\, PQ} \right) \nonumber \\
( {\sf{P}}^3_{\bf{1}} )_{IJK}^{\quad\;\, PQR} &=& \frac{1}{42} \, \Phi_{IJK} \Phi^{PQR} \nonumber \\
( {\sf{P}}^3_{\bf{7}} )_{IJK}^{\quad\;\, PQR} &=& \frac{1}{24} \, {*\Phi}_{IJKA} {*\Phi}^{PQRA} \nonumber \\
( {\sf{P}}^3_{\bf{27}} )_{IJK}^{\quad\;\, PQR} &=& \delta_{[I}^P \delta_J^Q \delta_{K]}^R - \frac{1}{42} \, \Phi_{IJK} \Phi^{PQR} -\frac{1}{24} \, {*\Phi}_{IJKA} {*\Phi}^{PQRA} \; .
\label{proj}
\eea
These can be checked using the $G_2$ identities in appendix A.

Smooth compact $G_2$ manifolds have a somewhat simpler cohomology due to the fact that all $H^I_{\bf{7}} =0$
(that is when the holonomy is the full $G_2$ and not a proper subgroup thereof).
The only independent non-trivial cohomology groups in this case are $H^3_{\bf{1}}$, $H^3_{\bf{27}}$ and $H^2_{\bf{14}}$.
A useful way to analyze the first two is to observe the isomorphism
\be
\alpha_{IJK} \; =\; 3 \Phi_{[IJ}^{\;\;\;\;\, A} \xi_{K]A} \; ,
\ee
between the components $\alpha_{IJK}$ in $\Lambda^3_{\bf{1}} \oplus \Lambda^3_{\bf{27}}$ and the symmetric tensor representation
$\xi_{IJ} = \xi_{JI}$ of $G_2$.
The traceless part $\xi_{IJ} - \frac{1}{7} \, g_{IJ} \xi^K_{\; K}$ of $\xi_{IJ}$ is isomorphic to $\Lambda^3_{\bf{27}}$ while its
trace part $\frac{1}{7} g_{IJ} \xi^K_{\; K}$ is isomorphic to the singlet representation $\Lambda^3_{\bf{1}}$.
Thus the only elements of $H^3_{\bf{1}}$ are constant multiples of $\Phi$. Furthermore one can show that if the 3-form $\alpha$
defined above is closed and coclosed (i.e. harmonic) then it follows that $\xi$ obeys
\be
\xi^I_{\; I} \; =\; 0 \; , \quad \nabla^I \xi_{IJ} \; =\; 0 \; , \quad \Phi_{I}^{\;\; KL} \nabla_K \xi_{LJ} \; =\; 0 \; .
\ee
The equations above are precisely those satisfied by the (linearly independent) small variations $\xi_{IJ} = \delta g_{IJ}$ of a
$G_2$ holonomy metric $g_{IJ}$ in order that the new metric $g_{IJ} + \delta g_{IJ}$ also has $G_2$ holonomy.
Thus elements of $H^3_{\bf{27}}$ correspond to such $G_2$ holonomy preserving deformations.
Finally, any element of $\Lambda^2_{\bf{14}}$ can be written as ${\sf{P}}^2_{\bf{14}} \beta$ for some 2-form $\beta$ on $M$.
Such elements have no other special properties, to the best of our knowledge, except that closure $d\, ({\sf{P}}^2_{\bf{14}} \beta) =0$ of
${\sf{P}}^2_{\bf{14}} \beta$ implies coclosure $d^\dagger ({\sf{P}}^2_{\bf{14}} \beta) =0$ identically.


\section{\large{Poincar\'{e} lemma for $\Lambda^2_{\bf 14}$}} \label{sec-pl}
\setcounter{equation}{0}

Given a 2-form ${\tilde B}$ on ${\mathbb{R}}^7$ in the ${\bf 14}$ irrep of $G_2$ that is coclosed $d^\dagger {\tilde B} =0$ then the Poincar\'{e} lemma can be used to deduce
\be
{\tilde B} \; =\; d^\dagger \Xi \; , \quad {\sf P}^2_{\bf 7} \, d^\dagger \Xi \; =\; 0\; ,
\label{poinlem}
\ee
for some 3-form $\Xi$. One can decompose $\Xi$ into irreps of $G_2$ as $\Lambda^3 = \Lambda^3_{\bf 1} \oplus \Lambda^3_{\bf 7} \oplus \Lambda^3_{\bf 27}$ using the 3-form projection operators in appendix B, such that
\be
\Xi_{MNP} \; =\; \Phi_{MNP} \, a + {*\Phi}_{MNPQ} \, b^Q + c_{MNP} \; ,
\ee
where $a={1\over 42} \Phi^{MNP} \, \Xi_{MNP}$, $b^Q = {1\over 24} {*\Phi}^{MNPQ} \, \Xi_{MNP}$ and $c \in \Lambda^3_{\bf 27}$.

The identity ${\sf P}^2_{\bf 14} d^\dagger {\sf P}^3_{\bf 1} =0$ together with ({\ref{poinlem}}) imply we can neglect $a$ in $\Xi$ because it will drop out of ${\tilde B} = {\sf P}^2_{\bf 14} {\tilde B} = {\sf P}^2_{\bf 14} d^\dagger \Xi$.

The identities in appendix A can be used to rewrite the second equation in ({\ref{poinlem}}) in components as
\be
\Phi_{MNP} \, \partial_Q \, \Xi^{NPQ} \; =\; -{4 \over 3}\, \Phi^{NPQ} \, \partial_{[N} \left( {*\Phi}_{PQM]R} \, b^R - c_{PQM]} \right) \; =\; 0 \; ,
\ee
where the identities $\Phi_{MIJ} \, c_{N}^{\;\;\;\, IJ} = \Phi_{NIJ} \, c_{M}^{\;\;\;\, IJ}$ and $\Phi^{MNP} c_{MNP} =0$ have also been used. Thus an equivalent form of this equation reads
\be
{\sf P}^4_{\bf 7} \, d \left( {\sf P}^3_{\bf 7} - {\sf P}^3_{\bf 27} \right) \Xi \; =\; 0 \; ,
\ee
a solution of which is
\be
\left( {\sf P}^3_{\bf 7} - {\sf P}^3_{\bf 27} \right) \Xi \; =\; d {\tilde \alpha} \; ,
\ee
where ${\tilde \alpha} \in \Lambda^2_{\bf 14}$. The reason that ${\tilde \alpha}$ is not a general 2-form is because of the identity ${\sf P}^3_{\bf 1} d {\sf P}^2_{\bf 14} =0$ implying $d {\tilde \alpha}$ is automatically in $\Lambda^3_{\bf 7} \oplus \Lambda^3_{\bf 27}$ as required by the equation.

Since we have assumed ${\sf P}^3_{\bf 1} \Xi =0$ then acting on the equation above with $( {\sf P}^3_{\bf 7} - {\sf P}^3_{\bf 27})$ gives
\be
\Xi \; =\; \left( {\sf P}^3_{\bf 7} - {\sf P}^3_{\bf 27} \right) d {\tilde \alpha} \; =\; \left( 2\, {\sf P}^3_{\bf 7} - 1 \right) d {\tilde \alpha}  \; ,
\ee
and hence
\be
{\tilde B} \; =\; d^\dagger \left( 2\, {\sf P}^3_{\bf 7} - 1 \right) d {\tilde \alpha} \; .
\ee
%


\section{\large{BV quantization of a free 6-form in 7 dimensions}} \label{sec-6form}
\setcounter{equation}{0}

The classical action for a free abelian $p$-form $\omega_p$ in $n$
dimensions is $S_0 = {1 \over 2} \int_n d \omega_p \w
{*d\omega_p}$. We will now describe the BV quantization of this
action for the special case of $p=6$ and $n=7$, where $\omega_6 =
C$.

The classical action for $\omega_6$ is degenerate under the gauge
transformations $\delta \omega_6 = d \lambda_5$. Thus we must
introduce a fermionic ghost $\omega_5$ for this symmetry. The
gauge symmetry is reducible for gauge parameters $\lambda_5 = d
\lambda_4$. This necessitates the addition of a ghost-for-ghost
bosonic field $\omega_4$. Continuing this line of reasoning leads
to a tower of descendent $p$-form ghosts $\omega_p$, associated to
$\omega_6$, with $0 \leq p<6$ and Grassmann parity $(-1)^{p}$.
Thus we have the collection $\Phi = \{ \omega_p | p=0,...,6 \}$ of
fields+ghosts with associated BRST transformations
\be Q \, \omega_p \; =\; d \omega_{p-1} \; , \ee
such that $Q \omega_0 =0$. The corresponding set of
anti(fields+ghosts) are $\Phi^* = \{ \chi_{7-p} | p=0,...,6 \}$,
where $\chi_{7-p}$ is a $(7-p)$-form with Grassmann parity
$(-1)^{p+1}$. The master equation $Q \Phi = \delta S / \delta
\Phi^*$ then fixes the form $\int \Phi^* \wedge Q \Phi$ of the
minimal contribution to the classical action from these fields.
The minimal solution of the master equation therefore corresponds
to the action
\be S \; =\; S_0 + \sum_{p=0}^5 \int \chi_{6-p} \w d \omega_{p} \;
,\ee
from which one derives the BRST transformations
\be Q \, \chi_1 \; =\; {d*d} \omega_6 \; , \quad Q \, \chi_p \;
=\; d \chi_{p-1} \; \; (p=2,...,6)\; , \ee
for the antifields. The BRST transformation of $\chi_7$ can be an
arbitrary BRST-invariant function. These transformations are
indeed nilpotent and generate a global symmetry of $S$.

To fix all the residual gauge symmetries in a systematic way
requires the introduction of quite an elaborate set of
non-minimal fields. We will not need to get into the details of
their BRST structure but let us just note that the appropriate
gauge fermion here is
\bea \Psi &=& \sum_{k=1}^6 \int \gamma_{8-k} \w d^\dagger \omega_k
+ \sum_{k=2}^6 \int \gamma_{8-k} \w d \theta_{k-2} + \sum_{k=3}^6
\int \theta_{k-2} \w d^\dagger \alpha_{10-k} \nonumber
\\
&& +\sum_{k=4}^6 \int \alpha_{10-k} \w d \beta_{k-4} + \int
\beta_1 \w d^\dagger \varepsilon_7 + \beta_2 \w d^\dagger
\varepsilon_6 + \int \varepsilon_6 \w d \varphi_0 \; . \nonumber \\
\eea
The form degree and parity of all the non-minimal fields appearing
here should be implicit. Imposing the gauge fermion constraint
$\Phi^* = \delta \Psi / \delta \Phi$ and integrating out the
Lagrange multiplier fields in the non-minimal terms in the action
then leads to the following antifield constraints
\bea \chi_k &=& d^\dagger \gamma_{k+1} \;\; (k=1,...,6) \; , \quad \chi_7 \; =\; 0 \nonumber \\
\gamma^*_k &=& d^\dagger \omega_{k+1} + d \theta_{k-1} \; =\; 0
\;\; (k=0,...,5) \nonumber \\
\theta^*_k &=& d \gamma_{k-1} + d^\dagger \alpha_{k+1} \; =\; 0
\;\; (k=3,...,7) \nonumber \\
\alpha^*_k &=& d^\dagger \theta_{k+1} + d \beta_{k-1} \; =\; 0
\;\; (k=0,1,2,3) \nonumber \\
\beta^*_k &=& d \alpha_{k-1} + d^\dagger \varepsilon_{k+1} \; =\;
0 \;\; (k=5,6,7) \nonumber \\
\varepsilon^*_k &=& d^\dagger \beta_{k+1} + d \varphi_{k-1} \; =\;
0 \;\; (k=0,1) \nonumber \\
\varphi^*_7 &=& d \varepsilon_6 \; =\; 0 \; .
\eea
Solving these equations implies the non-minimal fields $(
\varphi_0 , \varepsilon_{6,7} , \beta_{0,1,2} , \alpha_{4,5,6,7} ,
\theta_{0,1,2,3,4} )$ are harmonic, $\gamma_{2,3,4,5,6,7}$ are
closed and $\omega_{1,2,3,4,5,6}$ are coclosed. The latter
condition corresponds to the expected gauge-fixing constraint for
$p$-forms.

Imposing these constraints in the non-minimal action solving the
master equation leads us to the gauge-fixed action
\be S \; =\; {1 \over 2} \int \omega_6 \w * \Delta \, \omega_6 +
\sum_{k=1}^6 \int \gamma_{k+1} \w \Delta \omega_{6-k} \; .\ee
One can also verify that the aforementioned constraints solve the
equation $\int \Phi^* \w \Phi = \sum_{k=1}^6 \chi_k \w
\omega_{7-k} =0$, defining the graded Lagrangian submanifold in
configuration space.

Using the techniques of Schwarz that were reviewed in section 4,
we are now ready to compute the partition function for the free
6-form. The resolvent for the classical action $S_0$ here has the
associated differential complex
\be \label{res6} 0 \longrightarrow \Lambda^0
\overset{d}{\longrightarrow} .\, .\, .\,
\overset{d}{\longrightarrow} \Lambda^6 \overset{d^\dagger
d}{\longrightarrow} \Lambda^6 \longrightarrow 0 \; , \ee
where $n=6$, $\Gamma_i = \Lambda^{6-i}$, $T_i = d_{6-i}$ and the
extension by $K = d^\dagger_7 d_6$ has been included (using the
notation of section 4). With these identifications, Schwarz's
formula (\ref{partf}) for the partition function reads
\be Z_6 \; =\; ({\mbox{det}}\, d^\dagger_7 d_6 )^{-1/2} \left| {
{\mbox{det}}\, ( d_5 ) {\mbox{det}}\, ( d_3 ) {\mbox{det}}\, ( d_1
) \over {\mbox{det}}\, ( d_4 ) {\mbox{det}}\, ( d_2 )
{\mbox{det}}\, ( d_0 ) } \right| \; . \ee
The leading term can be written in a similar form to the other
terms using the identities $| {\mbox{det}}\, d_6 | =
({\mbox{det}}\, ( d^\dagger_7 ) \, {\mbox{det}}\, ( d_6 ) )^{1/2}
= ({\mbox{det}}\, d^\dagger_7 d_6 )^{1/2}$. This allows us to
identify $Z_6$ as the reciprocal of the Ray-Singer torsion
$I_{RS}$ of the 7-manifold (see e.g. equation (2.21) in \cite{PW}
for explicit identification). For our purposes it will be more
convenient to write $Z_6$ in terms of determinants of Laplacian
operators $\Delta = d d^\dagger + d^\dagger d$. This can be easily
achieved using standard properties of determinants (see
\cite{PW,SC}) to give
\be \label{pf6form} Z_6 \; =\; { ({\mbox{det}}\, \Delta^5 ) \over
({\mbox{det}}\, \Delta^6 )^{1/2} } { ({\mbox{det}}\, \Delta^3
)^{2} \over ({\mbox{det}}\, \Delta^4 )^{3/2} } { ({\mbox{det}}\,
\Delta^1 )^3 \over ({\mbox{det}}\, \Delta^2 )^{5/2} } { 1 \over
({\mbox{det}}\, \Delta^0 )^{7/2} } \; . \ee
Superscripts $\Delta^p$ denote the action of $\Delta$ on
$\Lambda^p$. It is perhaps worth concluding with a comment on why
we might expect this somewhat novel result. Recall that the
Ray-Singer torsion is a topological invariant of a differentiable
manifold in odd dimensions and can be understood as the analytic
torsion of the de Rham complex of the manifold. We refer to the
result as novel since $Z_6$ corresponds to the analytic torsion of
the complex (\ref{res6}) and not the de Rham complex (despite the
fact they are identical up to the last term). Nonetheless, we may
still have expected a topological invariant given that we are
describing the special case of a free 6-form in 7 dimensions --
which has no local on-shell degrees of freedom. Indeed this result
generalizes to any classical action $S_0 = {1 \over 2} \int_n d
\omega_{n-1} \w {*d \omega_{n-1}}$ describing a free $(n-1)$-form
in $n$ dimensions.

We end this appendix by noting a nice relation between the
partition function for a free $p$-form gauge field $\omega$ and a
free $(n-p-2)$-form gauge field ${\tilde \omega}$, in odd
dimensions $n=2k+1$, involving the Ray-Singer torsion. Recall that
such gauge fields describe equivalent local degrees of freedom in
that the field equation $d^\dagger G =0$ and Bianchi identity $dG
=0$ for $\omega$ (where $G=d\omega$) can also be written as $d
{\tilde G} = 0$, $d^\dagger {\tilde G} =0$, in terms of ${\tilde
G} = *G = d {\tilde \omega}$. Without loss of generality, we will
now assume $p=2r+1$ (the dual field will then always have even
degree in odd dimensions). The partition functions for $\omega$
and ${\tilde \omega}$ are
\bea Z_\omega &=&  ({\mbox{det}}\, \Delta^{2r+1} )^{-1/2} ({\mbox{det}}\, \Delta^{2r} ) ({\mbox{det}}\, \Delta^{2r-1} )^{-3/2} ... ({\mbox{det}}\, \Delta^{0} )^{r+1} \\
[.1in] Z_{\tilde \omega} &=& ({\mbox{det}}\, \Delta^{2(k-r)-2}
)^{-1/2} ({\mbox{det}}\, \Delta^{2(k-r) -3} ) ({\mbox{det}}\,
\Delta^{2(k-r)-4} )^{-3/2} ... ({\mbox{det}}\, \Delta^{0}
)^{-(k-r-1)-1/2} \; . \nonumber \eea
Some algebra and use of Hodge duality ${\mbox{det}}\, \Delta^{p} =
{\mbox{det}}\, \Delta^{n-p}$ then implies that the ratio
\be Z_{\omega} / Z_{\tilde \omega} \; =\; \prod_{i=0}^k (
{\mbox{det}}\, \Delta^{i} )^{(-1)^i ( (k-i) +1/2 )} \; =\; (
I_{RS} )^{(-1)^{k+1}} \; .  \ee
%


\section{\large{BV quantization of ${\tilde E}$}} \label{sec-Eform}
\setcounter{equation}{0}

Following the discussion of resolvents in section 4, we identify $\Gamma_0 = \Lambda^4_{{\bf 27} \oplus {\bf 1}}$ in the action $\int_{M_0} d{\tilde E} \w * \left( {3 \over 2}\, {\sf P}^5_{\bf 7} - 1 \right) d {\tilde E}$ for ${\tilde E}$ in ({\ref{finalvar2}}).
For suitable normalization of ${\tilde E}$, the kinetic operator in this action is
\be
K \; =\; - d^\dagger {\sf M} d \; =\; \left[ \Delta^4_{\bf 27} - {\sf P}^4_{\bf 27} d d^\dagger {\sf P}^4_{\bf 27} - {9 \over 2}\, {\sf P}^4_{\bf 1} d d^\dagger {\sf P}^4_{\bf 27} \right] - {1\over 2} \left[ \Delta^4_{\bf 1} - d d^\dagger {\sf P}^4_{\bf 1} \right]   \; ,
\ee
where ${\sf M} = {1 \over 2} {\sf P}^5_{\bf 7} - {\sf P}^5_{\bf 14}$. This kinetic operator is self-adjoint and indeed maps $\Lambda^4_{{\bf 27} \oplus {\bf 1}} \rightarrow \Lambda^4_{{\bf 27} \oplus {\bf 1}}$, which follows from the identity ${\sf P}^4_{\bf 7} d^\dagger {\sf M} d {\sf P}^4_{{\bf 27} \oplus {\bf 1}} =0$ using ${\sf P}^4_{\bf 7} d d^\dagger {\sf P}^4_{\bf 1} =0$.

The classical action for ${\tilde E}$ above is invariant under the gauge transformation $\delta {\tilde E} = {\sf P}^4_{{\bf 27} \oplus {\bf 1}} d \nu$, for any $\nu \in \Lambda^3_{{\bf 27} \oplus {\bf 7}}$ (the singlet part of $\nu$ is projected out of the gauge transformation). Furthermore this gauge symmetry is reducible for $\nu = {\sf P}^3_{{\bf 27} \oplus {\bf 7}} d \varepsilon$, for any 2-form $\varepsilon$. The projection operators do not commute with the exterior derivative so this statement is not obvious, but follows by noting ${\sf P}^3_{{\bf 27} \oplus {\bf 7}} d \varepsilon = d {\sf P}^2_{\bf 14} \varepsilon + {\sf P}^3_{{\bf 27} \oplus {\bf 7}} d {\sf P}^2_{\bf 7} \varepsilon$ and using that $d {\sf P}^3_{\bf 1} d {\sf P}^2_{\bf 7} \varepsilon \in \Lambda^4_{\bf 7}$. The remaining reducibilities are for $\varepsilon = d \xi$, for any 1-form $\xi$, and $\xi = d \gamma$, for any scalar $\gamma$.

In the notation of section 4, we therefore have a resolvent with $n=4$ and $T_1 = {\sf P}^4_{{\bf 27} \oplus {\bf 1}} d_3$, $T_2 = {\sf P}^3_{{\bf 27} \oplus {\bf 7}} d_2$, $T_3 = d_1$, $T_4 = d_0$ (their adjoints being just $T^\dagger = d^\dagger$
\footnote{Actually $T^\dagger_1 = {\sf P}^3_{{\bf 27} \oplus {\bf 7}} d^\dagger_4$ but this is identical to $d^\dagger_4$ when acting on elements of $\Lambda^4_{{\bf 27} \oplus {\bf 1}}$ since ${\sf P}^3_{\bf 1} d^\dagger {\sf P}^4_{{\bf 27}\oplus {\bf 1}} =0$.}
). The appropriate complex is
\be
0 \longrightarrow \Lambda^0 \overset{d}{\longrightarrow} \Lambda^1 \overset{d}{\longrightarrow} \Lambda^2 \overset{{\sf P}^3_{{\bf 27} \oplus {\bf 7}} d}{\longrightarrow} \Lambda^3_{{\bf 27} \oplus {\bf 7}} \overset{{\sf P}^4_{{\bf 27} \oplus {\bf 1}} d}{\longrightarrow} \Lambda^4_{{\bf 27} \oplus {\bf 1}}
\overset{K}{\longrightarrow} \Lambda^4_{{\bf 27} \oplus {\bf 1}} \longrightarrow 0 \; ,
\label{geng2res}
\ee
with the extension by $K$ included.

Using these identifications, the partition function
({\ref{partf}}) for $\int_{M_0} d{\tilde E} \w * \left( {3 \over 2}\, {\sf P}^5_{\bf 7} - 1 \right) d {\tilde E}$ can be written
\bea Z_4^{{\bf 27} \oplus {\bf 1}} &=& \left({\mbox{det}}\, K \right)^{-1/2} |{\mbox{det}}\, ( {\sf P}^4_{{\bf 27} \oplus {\bf 1}} d_3 ) | |{\mbox{det}}\, ( {\sf P}^3_{{\bf 27} \oplus {\bf 7}} d_2 ) |^{-1} |{\mbox{det}}\, ( d_1 ) | |{\mbox{det}}\, ( d_0 ) |^{-1} \nonumber \\ [.1in]
&=& ({\mbox{det}}\, \Delta^4_{{\bf 27} \oplus {\bf 1}} )^{-1/2} ({\mbox{det}}\, \Delta^3_{{\bf 27} \oplus {\bf 7}} ) ({\mbox{det}}\, \Delta^2 )^{-3/2} ({\mbox{det}}\, \Delta^1  )^{2} ({\mbox{det}}\, \Delta^0 )^{-5/2} \nonumber \\ [.1in]
&=& ( \det \, \Delta_{\bf 27} )^{1/2} ( \det \, \Delta_{\bf 14} )^{-3/2} ( \det \, \Delta_{\bf 7} )^{3/2} ( \det \, \Delta_{\bf 1} )^{-3} \; =\; I_{RS}^{-1}  \; .
\label{genpartf2}
\eea
The second equality has been obtained using
${\mbox{det}}\, ( K + T_1 T_1 ^\dagger ) = ( {\mbox{det}}\, K) |
{\mbox{det}}\, T_1 |^{2}$ and similar descendent identities for the resolvent.
The final equality uses the various $G_2$ isomorphisms described in section 5.4.
Thus we conclude that the partition function for ${\tilde E}$ is also equal to the inverse Ray-Singer torsion.


\section{\large{Hamiltonian action on $G_2$ string states}}
\setcounter{equation}{0}

In order to determine the action of Hamiltonian operators $H_L$
and $H_R$ on the spectrum of the $G_2$ string, the two main
examples to consider are states of the form $A_M (X) \psi_L^M$ in
${\bf 7} \otimes {\bf 1} \cong {\bf 7}$ and $B_{MN}(X) \psi_L^M
\psi_R^N$ in ${\bf 7} \otimes {\bf 7} \cong {\bf 14} \oplus {\bf
7} \oplus {\bf 27} \oplus {\bf 1}$. Note that the latter state
need have no definite (anti)symmetry properties since $\psi_L$ and
$\psi_R$ live in different sectors of the worldsheet theory.
Following appendix B, the symmetric part ${\bf 7} \otimes_s {\bf
7}$ is isomorphic to $\Lambda^3_{{\bf 27} \oplus {\bf 1}}$ while
the antisymmetric part ${\bf 7} \otimes_a {\bf 7}$ corresponds to
a general 2-form in $\Lambda^2_{{\bf 14} \oplus {\bf 7}}$. All
other cases can be mapped into these two examples this using the
isomorphisms between the various $G_2$ irreps in the exterior
algebra.

We start with the ${\bf 7} \otimes {\bf 7}$ case, since the other example follows easily from this one. We know from \cite{g2string} that the action of $Q_L$ on this state is given by
\be
Q_L \; : \; B_{M_1 N_1}  \; \longrightarrow \; ({\sf P}^2_{\bf 7})_{M_1 M_2}^{\;\; M_3 M_4} \nabla_{M_3} B_{M_4 N_1} \; ,
\ee
i.e. ${\check D}$ acting only on the left indices. The state $Q_L B$ is an element of $\Lambda^2_{\bf 7} \otimes \Lambda^1_{\bf 7}$. To define $H_L$, we also need to understand the action of the adjoint operator $Q_L^\dagger$. With respect to the standard inner product $\langle \omega , \xi \rangle = \int d^7 x \sqrt{g} \, g^{A_1 B_1} ... g^{A_n B_n} \, \omega_{A_1 ... A_n} \xi_{B_1 ... B_n}$ of rank $n$ tensors, the adjoint of $Q_L$ acting on $B$ is defined
\be
\langle \Omega, Q_L B  \rangle \; =\; \langle Q_L^\dagger \Omega, B \rangle \; ,
\ee
for any $\Omega \in \Lambda^2_{\bf 7} \otimes \Lambda^1_{\bf 7}$. The left hand side of this equation is given by
\begin{eqnarray*}
\langle \Omega , Q_L B \rangle &=& 6 \int d^7 x \sqrt{g} \, \Omega^{M_1 M_2 N_1} ({\sf P}^2_{\bf 7})_{M_1 M_2}^{\;\; M_3 M_4} \nabla_{M_3} B_{M_4 N_1} \\
&=& - \, 6 \int d^7 x \sqrt{g} \, \left( \nabla_{M_3} \Omega^{M_1 M_2 N_1} ({\sf P}^2_{\bf 7})_{M_1 M_2}^{\;\; M_3 M_4} \right)  B_{M_4 N_1} \\
&=& \langle Q_L^\dagger \Omega, B \rangle \; ,
\end{eqnarray*}
from which one reads off
\footnote{The extra factor of 6 comes via the identity $\phi_{MAB} \phi^{NAB} = 6 \delta_M^N$, which leads to the different normalizations of the $\Lambda^2_{\bf 7} \otimes \Lambda^1_{\bf 7}$ and $\Lambda^1_{\bf 7} \otimes \Lambda^1_{\bf 7}$ inner products.}
\be
Q_L^\dagger \; : \; \Omega_{M_1 M_2 N_1} \longrightarrow 6 \, \nabla^{M_2} \Omega_{M_3 M_4 N_1} ({\sf P}^2_{\bf 7})^{\;\; M_3 M_4}_{M_1 M_2} \; .
\ee

Thus we can now compute
\begin{eqnarray*}
(Q_L^\dagger Q_L B)_{M_1 N_1} &=& 6\, \nabla^{M_2} [ ({\sf P}^2_{\bf 7})_{M_3 M_4}^{\;\; M_5 M_6} \nabla_{M_5} B_{M_6 N_1} ] ({\sf P}^2_{\bf 7})^{\;\; M_3 M_4}_{M_1 M_2} \\
&=& 6\, \nabla^{M_2} \nabla_{M_3} B_{M_4 N_1} ({\sf P}^2_{\bf 7})^{\;\; M_3 M_4}_{M_1 M_2} \; ,
\end{eqnarray*}
using $( {\sf P}^2_{\bf 7} )^2 = {\sf P}^2_{\bf 7}$. Now substituting the explicit form $( {\sf{P}}^2_{\bf{7}} )_{IJ}^{\;\; PQ} = \frac{1}{3} \, \left( \delta_{[I}^P \delta_{J]}^{Q} + \frac{1}{2} \, {*\phi}_{IJ}^{\;\;\;\, PQ} \right)$ of the projector we find
\be
(Q_L^\dagger Q_L B)_{M_1 N_1} \; =\; - \nabla^2 B_{M_1 N_1} + \nabla^{M_2} \nabla_{M_1} B_{M_2 N_1} + \nabla^{M_2} \nabla_{M_3} B_{M_4 N_1} {*\phi}^{\;\;\;\, M_3 M_4}_{M_1 M_2} \; .
\ee
Notice that the right-sector index of $B$ has just gone along for the ride in the calculation above.

To get $H_L$ we still need to compute $Q_L Q_L^\dagger B$. It is easy to show that
\be
Q_L^\dagger \; :\;  B_{M_1 N_1} \longrightarrow - \nabla^{M_1} B_{M_1 N_1} \; ,
\ee
and then
\be
( Q_L Q_L^\dagger B)_{M_1 N_1} \; =\; - \nabla_{M_1} \nabla^{M_2} B_{M_2 N_1} \; .
\ee

Putting these results together gives
\bea
H_L \, B_{M_1 N_1} &=& -\nabla^2 B_{M_1 N_1} - [ \nabla_{M_1} , \nabla^{M_2} ] B_{M_2 N_1} + [ \nabla^{M_2} , \nabla_{M_3} ] B_{M_4 N_1} {*\phi}^{\;\;\;\, M_3 M_4}_{M_1 M_2}  \nonumber \\ [.1in]
&=& -\nabla^2 B_{M_1 N_1} - R_{M_1 M_2 N_1 N_2} B^{M_2 N_2} + R_{M_2 M_3 N_1}^{\quad\quad\;\;\;\; N_2}  B_{M_4 N_2} {*\phi}^{\;\;\, M_2 M_3 M_4}_{M_1}  \nonumber \\ [.1in]
& = & -\nabla^2 B_{M_1 N_1} - 3 R_{M_1 M_2 N_1 N_2} B^{M_2 N_2}  \; , \label{77lap}
\eea
where we have used $\phi^{AMN} R_{MNPQ} =0$ on a $G_2$ manifold (i.e. the curvature 2-form must transform in the adjoint ${\bf 14}$ of $G_2$). Consequently one finds that ${*\phi}_{MN}^{\;\;\;\, AB} R_{ABPQ} = -2\, R_{MNPQ}$ whose trace implies Ricci-flatness $R_{MN} =0$ by virtue of the Bianchi identity $R_{M[NPQ]} =0$. Both these results have also been used above.

Let us now decompose $B_{M_1 N_1}$ into symmetric and antisymmetric components and consider the action of $H_L$ on each component. If we take $B_{M_1 N_1}$ to be symmetric then ({\ref{77lap}}) corresponds to the Lichnerowicz Laplacian acting on a metric deformation of the $G_2$ manifold {\cite{g2string,Joyce}}. We can map the symmetric tensor $B$ to a 3-form $\omega$ in $\Lambda^3_{{\bf 27} \oplus {\bf 1}}$ via the isomorphism
\be
\omega_{IJK} \; =\; 3\,  \phi_{[IJ}^{\;\;\;\;\, A} B_{K]A} \; ,
\ee
described in appendix B. Thus, multiplying ({\ref{77lap}}) with $\phi$ followed by appropriate contraction and antisymmetrization, one obtains
\be
H_L \, \omega_{IJK} = - \nabla^2 \omega_{IJK} - {3 \over 2} R^{AB}_{\;\;\;\, [IJ} \omega_{K]AB} \; .
\ee
The expression above follows using the $G_2$ curvature identity $\phi^A_{\;\;\, [IJ} R_{K]ABC} = 0$ (this again follows from both pairs of indices of the Riemann tensor being in the {\bf 14} irrep of $G_2$).

Recalling the Weitzenboch formula
\be
(\Delta^p \omega)_{I_1 ...  I_p} = - \nabla^2 \omega_{I_1 ..  I_p} - {p \over (p-1)!}  R_{A[I_1}\omega^A{}_{I_2 ...  I_p]}
-{1 \over 4}{p(p-1) \over (p-2)!} R_{AB[I_1 I_2} \omega^{AB}{}_{I_3 \dots I_p]}
\ee
for $p$-forms we see that, under the map from symmetric tensors to
3-forms, $H_L$ maps to the ordinary 3-form Laplacian $\Delta^3 = d
d^\dagger + d^\dagger d$ on a $G_2$ manifold.

On the other hand, if $B$ is antisymmetric then one can easily
check that ({\ref{77lap}}) just reduces to the Weitzenboch formula
for 2-forms. Putting this together we have shown that $H_L = H_R =
\Delta^2 + \Delta^3_{{\bf 27} \oplus {\bf 1}}$ on states in the
${\bf 7} \otimes {\bf 7}$ representation. It will be convenient
to sometimes refer to this as $\Delta_{{\bf 7} \otimes {\bf 7}} =
\Delta^2_{\bf 14} + \Delta^2_{\bf 7} + \Delta^3_{\bf 27} +
\Delta^3_{\bf 1}$.

We now wish to determine the action of $H_L + H_R$ on states of the form $A_M (X) \psi_L^M$ in {\bf 7}$\otimes${\bf 1}. Repeating the calculation above for this simpler case we find
\bea
\{Q_L, Q_L^\dag \} A_M &=& - {1 \over 3} \left( \nabla^2 A_M - {1 \over 2} *\phi_M^{\;\;\; ABC} R_{ABCD} A^D \right) \; =\; -\frac{1}{3} \nabla^2 A_M \nonumber \\
\{Q_R, Q_R^\dag\}  A_M &=& -\nabla^2 A_M \; ,
\eea
using again Ricci-flatness $R_{MN} =0$ and the Bianchi identity $R_{M[NPQ]} =0$. Thus $H_L = H_R = \Delta^1 = \Delta^1_{\bf 7}$, up to normalization.


\end{document}